\documentclass[letterpaper, 10pt, journal, onecolumn, final]{IEEEtran}

\usepackage{graphics} 
\usepackage{epsfig} 
\usepackage{times}
\usepackage{soul}
\usepackage{comment}
\usepackage[normalem]{ulem}
\usepackage{bookmark}
\usepackage{booktabs}
\usepackage[utf8]{inputenc}
\usepackage[english]{babel}
\usepackage{mathtools}
\usepackage{amsfonts}
\usepackage{amsthm}
\usepackage{amssymb}
\usepackage{bbm}
\usepackage{color}
\usepackage{algorithm}
\usepackage[]{algpseudocode}
\usepackage{tikz}
\usepackage{cancel}
\usepackage{subcaption} 
\usepackage{ulem}
\usepackage{xparse}
\usepackage{xcolor}
\usepackage{todonotes}

\ExplSyntaxOn
\prop_new:N \g_author_color_prop
\prop_put:Nnn \g_author_color_prop {AT} {violet!30}
\prop_put:Nnn \g_author_color_prop {SB} {orange!30}
\prop_put:Nnn \g_author_color_prop {GN} {blue!30}

\NewDocumentCommand{\mytodo}{O{} m m}{%
  \prop_get:NnNTF \g_author_color_prop {#2} \l_tmpa_tl
    {\todo[inline,author=#2,color=\l_tmpa_tl,#1]{#3}}
    {\todo[inline,author=#2,#1]{#3}} %
}
\ExplSyntaxOff

\newcommand\oprocendsymbol{\hbox{$\square$}}
\newcommand\oprocend{\relax\ifmmode\else\unskip\hfill\fi\oprocendsymbol}

\renewcommand{\theenumi}{\arabic{enumi}.}

\newtheorem{theorem}{Theorem}[section]

\newtheorem{definition}[theorem]{Definition}
\newtheorem{lemma}[theorem]{Lemma}
\newtheorem{remark}[theorem]{Remark}

\makeatletter
\newcommand{\StatexIndent}[1][3]{%
  \setlength\@tempdima{\algorithmicindent}%
  \Statex\hskip\dimexpr#1\@tempdima\relax}
\makeatother

\newcommand{\real}{{\mathbb{R}}}
\newcommand{\integer}{{\mathbb{Z}}}
\renewcommand{\natural}{\mathbb{N}}

\newcommand{\fromto}[2]{\{#1,\ldots,#2\}}

\newcommand{\diag}{\operatorname{diag}}

\DeclarePairedDelimiter\norm{\lVert}{\rVert}

\newcommand{\infnorm}[1]{\left\| #1 \right\|_{\infty}}

\newcommand{\sml}{\scriptscriptstyle}
\newcommand{\metr}[1]{d_{\sml #1}}

\newcommand{\xdim}{n_x}
\newcommand{\ydim}{n_y}

\newcommand{\labelSet}{L}
\newcommand{\edges}{\mathcal{E}}
\newcommand{\adjConf}{\Sigma_{a}}
\newcommand{\adjConfp}{\adjConf^{\sml +}}
\newcommand{\adjConfn}{\adjConf^{\sml -}}
\newcommand{\adjConfInt}{\adjConf^\textit{int}}
\newcommand{\adjConfFront}{\adjConf^\textit{front}}
\newcommand{\adjConfFrontp}{\adjConf^{f \sml +}}
\newcommand{\numeq}{N_{\sml U}}

\newcommand{\qmarks}[1]{``#1"} 

\newcommand{\iter}{k}

\newcommand{\hIter}{h}
\newcommand{\indexStart}{i}
\newcommand{\indexStop}{j}
\newcommand{\transIter}{s}
\newcommand{\partIter}{r}
\newcommand{\cumulIter}{\zeta_{\transIter}^{\partIter}}

\newcommand{\state}{\mathcal{S}}
\newcommand{\statek}{\state^\iter}
\newcommand{\grid}{G}

\newcommand{\gridPoint}{p}
\newcommand{\gridPointNeigh}{q}
\newcommand{\Label}{\sigma}
\newcommand{\LabelInit}{\Label^0}
\newcommand{\Labelp}[1]{\Label_{#1}}
\newcommand{\Labelk}{\Label^\iter}
\newcommand{\Labelkk}{\Label^{\iter + 1}}
\newcommand{\prob}[1]{P\left(#1\right)}

\newcommand{\diffset}{\Theta}

\newcommand{\Type}{\tau}
\newcommand{\Typel}[1]{\tau_{#1}}
\newcommand{\Hamiltonian}{H}
\newcommand{\E}{E}
\newcommand{\labels}{\ell}
\newcommand{\typeSet}{\Gamma}
\newcommand{\types}{\gamma}
\newcommand{\J}{J}
\newcommand{\HamJ}{\Hamiltonian_\J}
\newcommand{\energy}{\tilde{\J}}
\newcommand{\Xset}{\mathcal{X}}
\newcommand{\Yset}{\mathcal{Y}}

\newcommand{\Neighbour}{\mathcal{N}}
\newcommand{\Kecm}{K_{\scalebox{0.5}{\textit{ECM}}}}

\newcommand{\ecm}{f}
\newcommand{\ecmInit}{\ecm_{\textit{init}}}
\newcommand{\ecmk}{\ecm^{\iter}}
\newcommand{\ecmkk}{\ecm^{\iter + 1}}
\newcommand{\ecmp}[1]{\ecm_{#1}}
\newcommand{\ecmpk}[1]{\ecm^\iter_{#1}}
\newcommand{\ecmpkk}[1]{\ecmp{#1}^{\iter + 1}}
\newcommand{\ecmlim}{\bar{\ecm}}

\renewcommand{\d}{d}
\newcommand{\dmax}{D}
\newcommand{\enzime}{z}
\newcommand{\enzimeDecay}[1]{a_{#1}}
\newcommand{\kecm}{c^\ecm}
\newcommand{\kenzime}[1]{c^\enzime_{#1}}
\newcommand{\targetVol}{V_T}
\newcommand{\HamV}{\Hamiltonian_V}
\newcommand{\KroneckerDelta}{\delta}
\newcommand{\nearcells}{n_\d}

\newcommand{\clball}[1]{\bar{\mathcal{B}}(#1)}
\newcommand{\numCells}{N}
\newcommand{\A}{A}
\newcommand{\uniformProb}{\mathcal{U}}
\newcommand{\uniformConfig}{\mathcal{A}_\Label}

\newcommand{\DeltaHmax}{\Delta \Hamiltonian_{\max}}
\newcommand{\DeltaHmin}{\Delta \Hamiltonian_{\min}}
\newcommand{\probLBound}{\varrho}
\newcommand{\expect}[1]{{\mathbb{E}\left[#1\right]}}

\newcommand{\absTime}[1]{T_{#1}^{\scalebox{0.5}{\textit{ INV}}}}
\newcommand{\meanAbsTime}[2]{\bar{T}_{#1 \,\mid\, #2}^{\scalebox{0.5}{\textit{ INV}}}}
\newcommand{\latticeDim}{n}
\newcommand{\labelEmpty}{\labels_{0}}
\newcommand{\labelSingleCell}{\labels_{1}}
\newcommand{\typeECM}{\types_{\scalebox{0.5}{\textit{ECM}}}}
\newcommand{\typeTumor}{\types_{c}}
\newcommand{\Vol}{V}
\newcommand{\initVol}{\Vol_0}
\newcommand{\Jcc}{\J_{c,c}}
\newcommand{\Jec}{\J_{\scalebox{0.5}{\textit{ECM}},c}}
\newcommand{\maxTumorDepth}[1]{\varphi_{\textit{max}}\left(#1\right)}
\newcommand{\probAbsTime}[4]{\mathcal{P}_{#1 \,\mid\, #2}\left(#3 \mid #4\right)} 
\newcommand{\absSet}{\mathcal{A}}

\newcommand{\truncHorz}{K}
\newcommand{\truncError}{R_\truncHorz}

\graphicspath{{figs/}}

\title{\LARGE \bf
System-Theoretic Analysis of Extended Cellular Potts Models \\ for Tumor Invasion
}

\author{A. Tramaloni$^{1}$, S. Baroncini$^{1}$, S. Avnet$^{2}$, N. Baldini$^{2,3}$ and G. Notarstefano$^{1}$
\thanks{$^{1}$ Department of Electrical, Electronic and Information Engineering, University of Bologna, Bologna, Italy.
\texttt{\{andrea.tramaloni, s.baroncini, giuseppe.notarstefano\}@unibo.it}.}
\thanks{$^{2}$ Department of Biomedical and Neuromotor Sciences, University of Bologna, Bologna, Italy. \texttt{\{sofia.avnet3, nicola.baldini5\}@unibo.it}.}
\thanks{$^{3}$ Biomedical Science, Technologies, and Nanobiotechnology Lab, IRCCS Istituto Ortopedico Rizzoli, Bologna, Italy. \texttt{nicola.baldini@ior.it}.}
}

\begin{document}

\maketitle
\thispagestyle{empty}
\pagestyle{empty}

\begin{abstract}
    The Cellular Potts Model has emerged as a standard framework for modeling multicellular systems and metastatic progression. Yet, computational extensions incorporating cell movement within spatial gradients of Extracellular Matrix (ECM), an important element of the tissue microenvironment, lack theoretical foundations.
    This work provides a system-theoretic formulation and analysis of Extended Cellular Potts Models (ECPMs) that explicitly couple tumor dynamics with ECM degradation.
    We reformulate the ECPM as an interconnected dynamical system: 
    a stochastic process describing a network of locally interacting tumor cells
    coupled to a jump-linear system describing ECM evolution.
    By exploiting the sparse energy structure of the model, we characterize asymptotic ECM behavior, transition probabilities with uniform bounds, and global system convergence.
    In one-dimensional settings, we derive closed-form expressions for invasion time probabilities, revealing how mean invasion time depends on local ECM gradients, number of cells and their tendency to maintain a target volume.
    Numerical simulations validate theoretical predictions, establishing a principled foundation for analyzing haptotaxis-driven invasion in Potts-based models.
\end{abstract}


\section{Introduction} \label{sec:introduction} 
Understanding the complex mechanisms governing tumor invasion and progression is a crucial goal in cancer research. Mathematical modeling plays a pivotal role in replicating the underlying principles of tumor dynamics, offering a structured framework to test hypotheses and interpret experimental findings. One class of models, the Cellular Potts Model (CPM), has emerged as a versatile tool to describe multicellular systems, capturing spatially resolved cell behaviors and interactions across multiple scales. 
A tumor can be viewed as a complex system with local interactions, in which agents (cells) interact locally with their neighbors and with a shared environment, giving rise to emergent collective behaviors such as invasion.
Traditionally, CPMs depict tissues as ensembles of cells governed by energy-based rules that account for cell adhesion and volume constraints. This approach allows the emergent behavior of cellular assemblies to be studied as a consequence of local, cell-cell interactions.
However, real tumor environments are far more complex, shaped not only by intercellular adhesion but also by interactions between cells and their surrounding extracellular matrix (ECM). 
These interactions give rise to a key process in tumor invasion where cells move along gradients of ECM concentration or adhesion strength. 
Classical CPMs account only for the intercellular interaction network, and neglect the feedback between the cells and the evolving environment.
Building on the foundational analysis provided in~\cite{voss2012multi}, which discusses the theoretical properties of standard CPMs, our work focuses on a specific subclass of Extended Cellular Potts Models (ECPMs) that explicitly incorporates cell-ECM interactions to model haptotaxis~\cite{oudin2016tumor}. 
Within this framework, we extend classical CPM theory by leveraging system-theoretic tools to address more realistic tumor invasion models, thus offering new insights into the mechanisms by which cell-ECM coupling influences invasion dynamics.

\subsubsection*{Literature Review}

In the following, we present the literature review divided into three categories: (i) early theoretical studies on CPMs, (ii) applied CPMs in cancer biology and their extensions ECPMs, and (iii) related works in the systems-and-control community.

Early work~\cite{graner1992simulation} introduces CPMs to simulate cell sorting via differential adhesion and energy minimization principles. 
Subsequent mathematical studies analyze critical behavior, phase transitions, and Gibbs measures to establish equilibrium properties in Potts models~\cite{beffara2012self, duminil2017continuity, rozikov2022gibbs}. 
Rigorous analytical frameworks further characterize CPM dynamics—including ergodicity, convergence, and parameter sensitivity~\cite{voss2012multi}—and connected discrete stochastic rules to continuum PDE limits~\cite{alber2007continuous, dan2005solving}. 
Despite these advances, most theoretical analyses have focused on the statistical mechanics or continuum limits of CPMs without explicitly incorporating ECM dynamics or cell--environment feedback, which are central to haptotaxis-driven invasion.

In biological applications CPMs have successfully modeled morphogenesis~\cite{savill1997modelling}, cancer invasion driven by intercellular adhesion~\cite{turner2002intercellular}, and drug treatment failure dynamics~\cite{turner2004tamoxifen}. 
Recent extensions, i.e. ECPMs for tumor invasion, explicitly incorporate ECM interactions, modeling tumor invasion~\cite{szabo2012invasion}, angiogenesis driven by matrix gradients~\cite{daub2013cell}, dynamic ECM remodeling in wound healing~\cite{scianna2015extended}, and broader tissue organization~\cite{hirashima2017cellular, crossley2024modeling}. 
While these works provide powerful computational tools, they largely remain simulation-oriented, with limited theoretical formalization and results on the underlying dynamics.
In particular, a rigorous theoretical understanding of tumor-cells-ECM feedback within ECPMs is still missing.

From a systems-and-control perspective, several recent works have addressed related classes of stochastic and interconnected dynamical systems. 
Markov jump linear systems have received renewed attention in the analysis and design of stochastic control systems~\cite{janschporto2023policy}, and switched and positive linear system formulations have been employed to study analogous transition mechanisms over networks~\cite{ogura2017optimal}.
In parallel, a substantial body of work in the network-control community has focused on discrete-time Markov-chain models of spreading and diffusion processes over networks, which, like the ECPM, are stochastic dynamics driven by local interactions.
These works provided stability, convergence, and estimation results for epidemic and related processes~\cite{pare2018epidemic, pare2020analysis, ogura2016stability, lambiotte2015random}.
Control-theoretic tools have also been applied directly to cellular automata to study morphogenetic processes~\cite{beros2018morphogenetic}, and, more recently, to the coordination and control of biological cell
populations, ranging from microbial consortia to spatially organized multi-agent systems~\cite{martinelli2023multicellular, brancato2024ratiometric, giusti2023local}. 
Bio-inspired swarm models have similarly been proposed to describe cancer progression in control-theoretic terms~\cite{raz2021target}, while computational approaches from the systems-biology community have investigated how ECM manipulation affects local tumor invasion~\cite{pal2020manipulating}. 
These contributions reflect a growing interest in applying systems theory to biologically motivated stochastic and networked models, a direction that directly inspires our approach.

\subsubsection*{Contributions}

The contributions of this paper are as follows.
We introduce a novel system-theoretic framework for analyzing ECPMs that explicitly incorporate cell-ECM interactions responsible for haptotaxis-driven tumor invasion.
Unlike classical CPMs, typically explored through empirical simulations, we establish a rigorous analytical foundation for these models by reformulating the ECPM as a closed-loop dynamical system composed of two interconnected subsystems: (i) a stochastic process governing the tumor cell configuration and (ii) a jump-linear system describing ECM evolution. 
We derive analytical results characterizing key properties of the coupled system, including the asymptotic behavior of the ECM field, uniform bounds on transition probabilities of tumor configurations, and convergence properties of the global dynamics. These results provide the first rigorous theoretical description of haptotaxis within the CPM framework.
Moreover, in one-dimensional settings we exploit the sparse energy structure of the model to obtain closed-form expressions for invasion time probabilities. These expressions reveal how invasion time and stochastic propagation depend explicitly on haptotactic strength, ECM degradation rates, number of cells and their target volumes.
Numerical simulations validate the theoretical predictions, confirming the analytical results for estimation of mean invasion time. 
This work bridges computational and theoretical analysis of ECPMs, by establishing a principled foundation for analyzing ECM-driven tumor invasion in Potts-based models.
Finally, the theoretical tools for interconnected stochastic and jump-linear systems may pave the way to a broader class of control systems inspired by biology.

\subsubsection*{Paper Organization}
The paper is organized as follows.
Section~\ref{sec:standard_formalism} introduces the standard ECPM formalism, covering lattice structure, energy formulation, and the conventional algorithm used to simulate these models.
Section~\ref{sec:system_theoretical_viewpoint} reformulates the ECPM as an interconnected dynamical system coupling stochastic cell configuration dynamics with jump-linear ECM evolution, emphasizing the sparse energy structure. 
Section~\ref{sec:analysis_and_properties} analyzes system properties: asymptotic ECM behavior, transition probabilities and their bounds, and global convergence to absorbing states. 
Section~\ref{sec:1Dgrid_absorption_times} derives closed-form invasion time probability expressions for one-dimensional ECPMs. 
Section~\ref{sec:simulations} validates theoretical predictions through Monte Carlo simulations.

\paragraph*{Notation}
For any finite set $S$, $|S|$ denotes its cardinality. We use $\norm{\cdot}$ for the Euclidean norm. Any function $F:S\to\real$ is identified, up to isomorphism, with a vector in $\real^{|S|}$ and is denoted by the same symbol. We write $\KroneckerDelta(\cdot,\cdot)$ for the Kronecker delta, $\prob{\cdot}$ for probability, $\expect{\cdot}$ for expectation, and $\uniformProb(S)$ for the uniform distribution on $S$.
Throughout the paper we will use the terms \qmarks{grid} and \qmarks{lattice} interchangeably, with \qmarks{lattice} emphasizing the mathematical structure.

\section{Conventional Algorithmic Formulation of Extended Cellular Potts Models}\label{sec:standard_formalism}

In this section, we recall the standard ECPM algorithmic formulation for simulating ECM-driven tumor invasion and set up the notation for the system-theoretic formalization and analysis developed in the next sections.

In ECPMs, the spatial domain is discretized into a regular grid of sites, each representing a small patch of ECM that can be occupied by cells or remain empty. 
Specifically, let us consider a two-dimensional finite lattice $\grid := \Xset \times \Yset \subset \mathbb{Z}^2$ with dimensions $\xdim \coloneqq |\Xset|$ and $\ydim \coloneqq |\Yset|$.
On $\grid$ there exist boundary conditions consisting of zero-flux at the top and bottom boundaries, and periodic conditions along the left and right sides of the grid. 
Zero-flux at the top/bottom models confinement within the epithelial boundary layer of a lumen, preventing motion in that direction. 
Instead, the periodic lateral boundary conditions approximate a small portion of a much larger domain.
Let us introduce a finite set of labels $\labelSet \subset \natural$.
Then, each grid point $\gridPoint = \left(x_\gridPoint, y_\gridPoint\right) \in \grid$ is associated with a label through the map
\begin{equation}\label{eq:label_map}
    \begin{aligned}
        \Label : \grid &\to \labelSet \\
        \gridPoint &\mapsto \Labelp{\gridPoint}.
    \end{aligned}
\end{equation}
Here, each label $\Labelp{\gridPoint} \in \labelSet$ identifies the entity occupying the grid point $\gridPoint \in \grid$, which can belong to a tumor cell or empty space. Thus, the label map in~\eqref{eq:label_map} provides a complete representation of the spatial configuration of all entities on the grid.
Moreover, each label $\labels \in \labelSet$ is associated with a specific type via the following mapping 
\begin{equation}\label{eq:type_map}
    \begin{aligned}
        \Type : \labelSet &\to \typeSet \\ 
        \labels &\mapsto \Typel{\labels},
    \end{aligned}
\end{equation}
where $\typeSet \subset \natural$ is a finite set containing grid point types that represent either cell types (e.g., tumor cells or normal cells) or the ECM.
We now introduce the ECM density scalar field $\ecm$ over the discrete lattice.
Formally, 
\begin{equation}\label{eq:ecm_field}
    \begin{aligned}
        \ecm : \grid &\to \real_{\geq 0} \\ 
        \gridPoint &\mapsto \ecmp{\gridPoint},
    \end{aligned}
\end{equation}
that evolves over time due to degradation by proteolytic enzymes secreted by tumor cells.
In practice, the ECM field is uniformly initialized on the lattice to a constant value $\ecmInit \in \real_{>0}$, i.e. $\ecmp{\gridPoint} = \ecmInit$ for all $\gridPoint \in \grid$.
\begin{figure}[htpb!]
    \centering
    \includegraphics[scale=0.65]{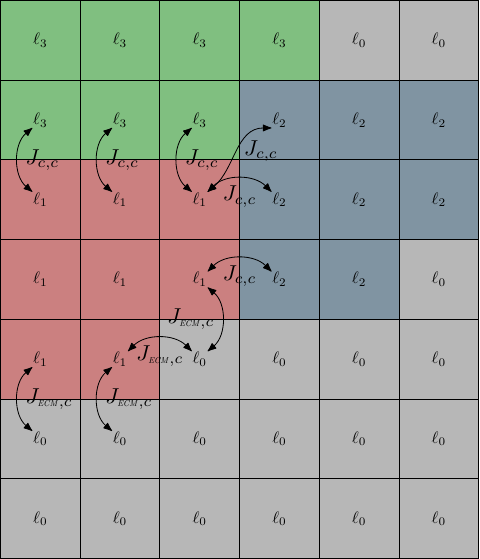}
    \caption{Cell-surface interactions in the ECPM are regulated by energy coefficients $\Jcc$ (cell-cell) and $\Jec$ (cell-ECM) adhesion. Here, a single tumor cell (red) interacts with neighboring cells (green and blue) and ECM (gray).}
    \label{fig:scheme}
\end{figure}
Before presenting the specific update rule of~\eqref{eq:ecm_field}, we introduce some quantities defined on the grid.  
Consider two arbitrary grid points $\gridPoint,q \in \grid$, then we equip the grid with the following Chebyshev distance metric
\begin{equation*}\label{eq:distance}
    \metr{\grid}(\gridPoint,q) := 
    \max\left(\varDelta_x\left(\gridPoint, q\right), \; \norm{y_{\gridPoint} - y_\gridPointNeigh}\right),
\end{equation*}
where $\varDelta_x\left(\gridPoint, q\right) := \min\left(\norm{x_{\gridPoint} - x_\gridPointNeigh}, \xdim - \norm{x_{\gridPoint} - x_\gridPointNeigh}\right)$ accounts for the periodic boundary conditions along the horizontal direction.
The metric $\metr{\grid}$ is the shortest-path distance of the undirected graph
$(\grid,\edges)$, with edges $\edges := \{(\gridPoint,\gridPointNeigh) \in \grid^2 :
\metr{\grid}(\gridPoint,\gridPointNeigh)=1\}$, whose nodes are the lattice sites.
Then, the closed ball of radius $\d \in \natural$ centered at grid point $\gridPoint$ can be expressed as $\clball{\gridPoint,\d} := \left\{\, q \in \grid : \metr{\grid}(\gridPoint,q) \le \d \,\right\}$.
For $\types\in\typeSet$ and $\d > 0$ we define the number of cells of type $\types$ within distance $\d$ from $\gridPoint$ by
\begin{equation*}\label{eq:nearcells}
    \nearcells(\gridPoint,\types,\Label) := \left|\left\{\labels \in \Label(\clball{\gridPoint,\d}) : \Typel{\labels} = \types \right\}\right|.
\end{equation*}
Next, we introduce the model parameters that will appear in the $\ecm$ dynamics.  
For each cell type $\types\in\typeSet$, let $\kenzime{\types} \geq 0$ be the enzyme secretion rate, $\enzimeDecay{\types} \geq 0$ the spatial decay coefficient, and $\dmax_\types\in\natural$ the maximal influence radius.  
Let $\kecm_\types \geq 0$ denote the ECM degradation rate induced by cells of type $\types$.
With these definitions in place, the proteolytic enzyme produced at $\gridPoint \in \grid$ by cells of type $\types$ is given by
\begin{equation}\label{eq:enzime}
    \enzime(\gridPoint,\types, \Label) := 
    \kenzime{\types} 
    \sum_{\d=1}^{\dmax_\types} 
    \nearcells(\gridPoint,\types,\Label)\,
    e^{-\enzimeDecay{\types}\d},
\end{equation}
from which follows the ECM update rule, expressed as 
\begin{equation}\label{eq:ecm}
    \ecmpkk{\gridPoint} = 
    \left(\,1 - 
    \sum_{\types\in\typeSet} 
    \kecm_\types\,\enzime(\gridPoint,\types, \Labelk)\,\right)
    \ecmpk{\gridPoint},
    \quad \forall \gridPoint \in \grid,
\end{equation}
where $\ecmpk{\gridPoint}$ and $\ecmpkk{\gridPoint}$ denote the ECM density at lattice point $\gridPoint$ at two consecutive iterations.
Note that the ECM update at each node depends only on the labels within graph
distance $\dmax_\types$, i.e., degradation is driven by the local neighborhood of each node.
To ensure physical consistency, we assume that the parameters $\kenzime{\types}$ and $\kecm_\types$ are chosen such that
\begin{equation}\label{eq:ecm_constraint}
    \sum_{\types\in\typeSet} 
    \kecm_\types\,\enzime(\gridPoint,\types,\Label)
    \;\le\; 1 \quad \forall \gridPoint \in \grid,
\end{equation}
namely, the ECM density cannot become negative after degradation.
Next, we define the interaction energy between neighboring grid points via the map $\J : \typeSet \times \typeSet \to \real_{\geq 0}$, which quantifies the interfacial energy between pairs of cell or medium types. 
Biologically, $\J$ captures the energetic cost of maintaining contact between two entities, reflecting effects such as cell--cell or cell--matrix adhesion and cortical tension.
In this work, we adopt the standard ECPM convention (see~\cite{turner2002intercellular,turner2004tamoxifen}), where the interaction energy is specified as
\begin{subequations}\label{eq:interaction_energy_standard}
    \begin{align}
        \J(\typeECM, \typeECM) &= 0 \label{subeq:J_ecm_ecm} \\
        \J(\typeTumor, \typeTumor) &= \Jcc \\
        \J(\typeTumor, \typeECM) &= \J(\typeECM, \typeTumor) = \Jec,
    \end{align}
\end{subequations}
where $\Jcc \in \real_{\geq 0}$ and $\Jec \in \real_{\geq 0}$ denote the cell-cell and cell-ECM adhesion energies, respectively. 
Here, $\typeECM \in \typeSet$ is the type corresponding to an empty grid point covered by ECM, and $\typeTumor \in \typeSet$ is the tumor cell type.
This reflects the assumption that ECM-ECM interfaces are energetically neutral, while cell-cell and cell-ECM contacts are penalized according to their adhesion strengths.
Consequently, the interaction energy between labels is defined as
\begin{equation}\label{eq:interaction_energy_func}
    \begin{split}
        \energy : \labelSet \times \labelSet &\to \real_{\geq 0} \\
        (\labels_1, \labels_2) &\mapsto \J(\Typel{\labels_1}, \Typel{\labels_2}) \left[1 - \KroneckerDelta\left(\labels_1, \labels_2\right)\right],
    \end{split}
\end{equation}
where the Kronecker delta ensures that only interfaces between different labels contribute to the energy. 
Figure~\ref{fig:scheme} illustrates these cell-surface interactions as regulated by $\energy$.
Define the set of neighbors of a point $\gridPoint$ as
$\Neighbour_\gridPoint := \{ \gridPointNeigh \in \grid : (\gridPoint,\gridPointNeigh) \in \edges \}$.
Hence, a cell is a set of nodes sharing the same label, and adhesion energy arises only along the edges connecting nodes with different labels.
Then, we introduce the core term in CPMs, i.e. the total energy of the system $\Hamiltonian: \labelSet^{|\grid|} \to \real_{\geq 0}$, represented by the Hamiltonian function $\Hamiltonian\left(\Label\right) := \HamJ\left(\Label\right) + \HamV\left(\Label\right)$, where
\begin{align}
    \HamJ\left(\Label\right) &:= \frac{1}{2} \sum_{\gridPoint \in \grid} \sum_{q \in \Neighbour_\gridPoint} \energy(\Labelp{\gridPoint}, \Labelp{\gridPointNeigh}) \label{subeq:HamJ}\\
    \HamV\left(\Label\right) &:= \lambda \sum_{\labels \in \labelSet \setminus \{\labelEmpty\}} \left(\sum_{w \in \grid}\KroneckerDelta(\Labelp{w}, \labels) - \targetVol\right)^2. \label{subeq:HamV}
\end{align}
Here, $\HamJ$ collects all local interaction contributions between neighboring sites, 
while $\HamV$ is the volume penalty term that favors label-wise occupancy close to a target value. 
The parameter $\lambda \in \real_{>0}$ weighs the strength of the volume constraint, thus regulating how strongly cells tend to maintain their target volume.
In~\eqref{subeq:HamV}, $\labelEmpty$ represents the label corresponding to empty lattice sites covered by ECM. 
The label $\labelEmpty$ is excluded from the sum since empty/ECM sites do not contribute to $\Hamiltonian$. 

ECPMs (and CPMs) are simulated in practice through a Markov chain Monte Carlo method, the so-called \emph{Modified Metropolis Algorithm} (MMA)~\cite{anderson2007single, binder2012monte}. 
The latter constitutes an adaptation of the classical \emph{Metropolis algorithm}~\cite{bhanot1988metropolis} in the context of Potts models. 
The MMA is described step-by-step in Algorithm~\ref{alg:MMA}.
\begin{algorithm}[htpb!]
\caption{MMA for simulating ECPM}
\label{alg:MMA}
\begin{algorithmic}[1]
    \State Start from an initial configuration $(\Label, \ecm)$
    \Repeat
        \State Select grid point $\gridPoint \sim \uniformProb(\grid)$ \label{step:select_p}
        \State Pick nearest neighbor $q \sim \uniformProb(\Neighbour_\gridPoint)$
        \If{$\Labelp{\gridPoint} = \Labelp{\gridPointNeigh}$}
            \State \textbf{goto} step \ref{step:select_p}
        \EndIf

        \State Define trial configuration $\Label'$ such that
        \begin{equation*}
            \Labelp{z}^\prime = \Labelp{z} + \left(\Labelp{\gridPoint} - \Labelp{\gridPointNeigh}\right) \KroneckerDelta(z,q) \; \forall z \in \grid
        \end{equation*}

        \State Evaluate energy change $\Delta\Hamiltonian = \Hamiltonian(\Label^\prime) - \Hamiltonian(\Label)$
        
        \State Compute total energy 
        \begin{equation}\label{step:DeltaE}
            \Delta\E(\Label, \ecm) = \Delta\Hamiltonian + \Kecm \left(\ecmp{\gridPoint} - \ecmp{\gridPointNeigh}\right)
        \end{equation} 

        \If{$\Delta\E(\Label, \ecm) < 0$} \label{step:accept_favorable}
            \State Accept the copy and update $\Labelp{\gridPointNeigh} \gets \Labelp{\gridPoint}$
        \Else
            \State \label{step:accept_unfavorable} Accept the change $\Labelp{\gridPointNeigh} \gets \Labelp{\gridPoint}$ with probability
            \begin{equation}\label{step:acceptance_prob}
                e^{-\frac{\Delta\E(\Label, \ecm)}{\beta}}
            \end{equation}
        \EndIf
        \State Update $\ecm_z \; \forall z \in \grid$ via dynamics \eqref{eq:ecm}
    \Until{stopping criterion}
\end{algorithmic}
\end{algorithm}
In equations~\eqref{step:DeltaE} and~\eqref{step:acceptance_prob}, $\Kecm \geq 0$ is a model parameter that weighs the influence of ECM gradients on cell migration, while parameter $\beta > 0$ influences the stochastic behavior of the dynamics, in fact it is usually related to environmental noise.
Note that each MMA iteration is a local asynchronous update, depending only on the neighborhood of the selected node.
\begin{figure}[htpb!]
    \centering
    \includegraphics[width=0.55\columnwidth]{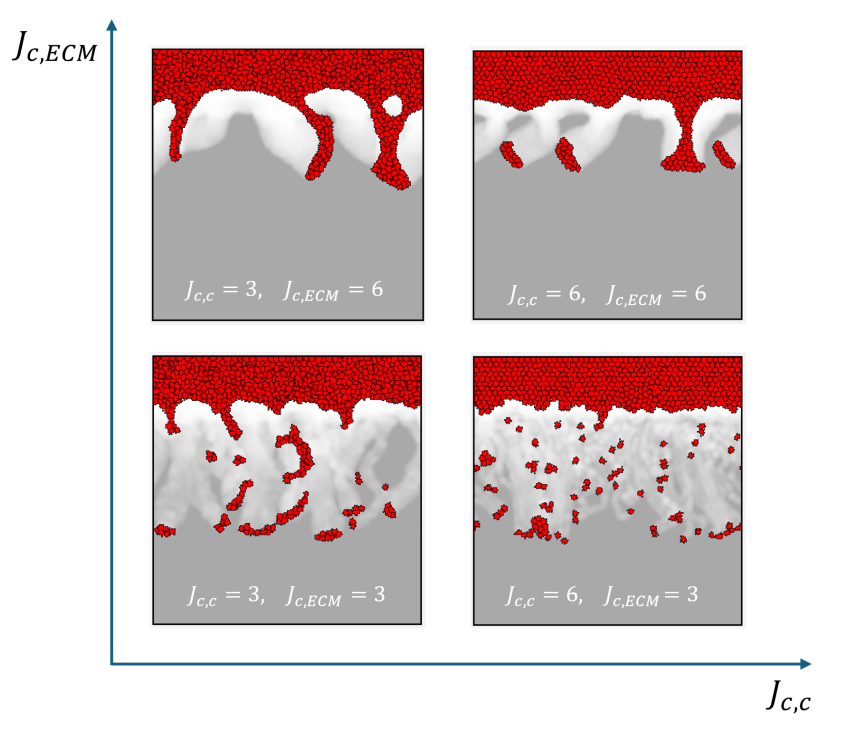}
    \caption{ECPM simulation on a 2D lattice showing tumor invasion (red) through ECM (gray) at different adhesion energies $\Jcc$ (cell-cell) and $\Jec$ (cell-ECM).}
    \label{fig:ecpm_standard_simulation}
\end{figure}
While the algorithmic description enables simulation of ECPM dynamics, it does not provide analytical insight into the underlying system properties. 
As an example, Figure~\ref{fig:ecpm_standard_simulation} shows a simulation on a 2D lattice, where varying the adhesion energies $\Jcc$ and $\Jec$ produces different invasion patterns. 
However, such simulations are descriptive and do not support a systematic analysis of the model's behavior. 
To overcome this limitation, the next section reformulates the ECPM in a system-theoretic framework, enabling a principled study of its dynamics.

\section{ECPM as Sparse Interconnected Systems} \label{sec:system_theoretical_viewpoint}
In this section, rather than treating the ECPM as an algorithmic rule, we regard it as an interconnected dynamical system composed of two distinct yet coupled subsystems: 
(i) a stochastic process describing a network of locally interacting tumor cells, influenced by the ECM substrate, and (ii) a jump-linear system governing the ECM density field, whose transition matrix changes due to the current tumor mass configuration.
\begin{figure}[htpb!]
    \centering
    \includegraphics[width=0.3\columnwidth]{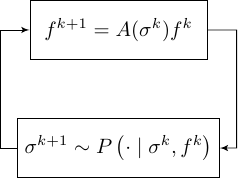}
    \caption{System-theoretic view of the ECPM. The discrete stochastic process is coupled to a jump-linear system.}
    \label{fig:interconnection}
\end{figure}
We start by introducing the ECM transition matrix underlying dynamics~\eqref{eq:ecm} as follows
\begin{equation}\label{eq:trans_matrix}
    \A(\Label) := \diag_{\gridPoint \in \grid} \left\{\A_\gridPoint(\Label)\right\},
\end{equation}
with the local terms defined as
\begin{equation}\label{eq:trans_matrix_local}
    \A_\gridPoint(\Label) := 1 - \alpha_\gridPoint(\Label) = 1 - \sum_{\types \in \typeSet} \kecm_\types \enzime(\gridPoint,\types,\Label),
\end{equation}
where $\alpha_\gridPoint(\Label)$ represents the total degradation rate at site $\gridPoint$ induced by all cell types present at current configuration $\Label$.
Here, $\A_\gridPoint(\Label) \in \left[0, 1\right]$ since $0 \leq \alpha_\gridPoint(\Label) \leq 1$ due to condition~\eqref{eq:ecm_constraint} and being the parameters within~\eqref{eq:enzime} non-negative.
In view of the definition in~\eqref{eq:ecm}, the ECM field can be identified with a vector $\ecm \in \real^{|\grid|}_{\geq 0}$ (see \emph{Notation}). 
Accordingly, its dynamics can be rewritten as the following jump-linear system
\begin{equation}\label{eq:ecm_jump_linear}
    \ecmkk = \A(\Labelk) \ecmk,
\end{equation}
with $\Labelk$ and $\ecmk$ being the cell mass configuration and ECM field on the lattice at discrete-time (iteration) $\iter \in \natural$, respectively.
Moreover, since the acceptance probability in~\eqref{step:acceptance_prob} depends on the local ECM values $\ecmp{\gridPoint}$ and $\ecmp{\gridPointNeigh}$ (see $\Delta \E$ definition at MMA step~\ref{step:DeltaE}), the $\Label$-dynamics are thus influenced by $\ecm$. 
Formally, the stochastic process associated with the evolution of $\Label$ can be generally written as
\begin{equation}\label{eq:tumor_stochastic_process}
    \Labelkk \sim \prob{\cdot \mid \Labelk, \ecmk},
\end{equation}
where the transition probability is given by
\begin{equation}\label{eq:transition_prob}
    \prob{\cdot \mid \Label, \ecm} :=
    \begin{cases}
        1 \quad & \text{if } \Delta \E(\Label, \ecm) \leq 0 \\
        e^{-\frac{\Delta \E(\Label, \ecm)}{\beta}} \quad & \text{otherwise} 
    \end{cases},
\end{equation}
that is a piecewise Boltzmann-weighted probability derived from the MMA steps~\ref{step:accept_favorable}-\ref{step:accept_unfavorable} (see Algorithm~\ref{alg:MMA}).
The resulting feedback interconnection between the subsystems~\eqref{eq:ecm_jump_linear} and~\eqref{eq:tumor_stochastic_process} is pictured in Fig.~\ref{fig:interconnection}.
Based on this reformulation, we analyze the Hamiltonian variation $\Delta \Hamiltonian$. 
Local cell-cell and cell-ECM interactions produce a sparse energy structure determined by the interaction graph $(\grid, \edges)$ as per~\eqref{subeq:HamJ}. 
This sparsity links the stochastic process~\eqref{eq:tumor_stochastic_process} of interacting tumor cells directly to local energy terms, enhancing analytical tractability.
This property underpins the derivation of closed-form expressions for invasion time probabilities presented in Section~\ref{sec:1Dgrid_absorption_times}.
We formalize this sparsity property in the following lemma.

\begin{lemma}[Sparse Energy Decomposition]\label{lem:energy_change_decomposition}
    Consider an ECPM at current state $\left(\Label, \ecm\right)$. 
    Let $\Label'$ be an adjacent configuration that differs from $\Label$ only at neighboring lattice sites $\gridPoint \in \grid$ and $q \in \Neighbour_{\gridPoint}$.
    Then, the total energy change $\Delta \E(\Label, \ecm)$ can be decomposed into sparse energy terms as follows
    \begin{equation*}
        \Delta \E(\Label, \ecm) = \Delta \HamJ(\Label) + \Delta \HamV(\Label) + \Kecm\left(\ecmp{\gridPoint} - \ecmp{\gridPointNeigh}\right),
    \end{equation*}
    where the adhesion-energy change is given by
    \begin{equation}\label{eq:deltaHJ_simplified}
        \Delta \HamJ(\Label) = \sum_{w \in \Neighbour_{q}} \energy(\Labelp{\gridPoint}, \Labelp{w}) - \energy(\Labelp{\gridPointNeigh}, \Labelp{w}),
    \end{equation}
    and the volume-energy change simplifies to
    \begin{equation}\label{eq:deltaHV_simplified}
        \Delta \HamV(\Label) = \lambda \!\sum_{\labels \in \{\Labelp{\gridPoint},\Labelp{\gridPointNeigh}\}} \!\left[1 + 2\rho(\Labelp{\gridPoint},\labels) (\Vol_\labels - \targetVol)\right] (1-\KroneckerDelta(\labelEmpty,\labels)),
    \end{equation}
    with 
    \begin{equation*}
        \Vol_\labels := \sum_{w \in \grid} \KroneckerDelta(\Labelp{w},\labels), \quad \rho(\labels_1,\labels_2) := 2\KroneckerDelta(\labels_1,\labels_2)-1 \in \{-1,1\}.
    \end{equation*}
\end{lemma}

\begin{proof}
    See Appendix~\ref{app:proof_lem_energy_decomposition}.
\end{proof}

This result highlights how the total energy change decomposes into three sparse terms that determine the stochastic evolution in~\eqref{eq:tumor_stochastic_process} following the probability rule in~\eqref{eq:transition_prob}.
Therefore, the state dynamics of the ECPM can be interpreted as a stochastic process driven by a sparse energy functional, coupled to a jump-linear system representing the ECM degradation. 
This novel interpretation clarifies the feedback mechanisms between tumor cells and the ECM, making the interaction dynamics in this class of models more explicit and analytically tractable.

\section{System Analysis and Properties} \label{sec:analysis_and_properties}
In this section we analyze the dynamical properties of the ECPM as formulated in Section~\ref{sec:system_theoretical_viewpoint}. 
We begin by presenting a result on the asymptotic behavior of the ECM field $\ecm$, and then we characterize the transition probabilities related to the $\Label$-dynamics. 
Then, we define uniform lower and upper bounds on the transition probabilities as a function of model parameters. 
Finally, we fully characterize the global system behavior and its convergence properties.
In the following, we denote by $\statek = (\Labelk, \ecmk)$ the global ECPM state at iteration $\iter$, with dynamics as in~\eqref{eq:ecm_jump_linear}-\eqref{eq:tumor_stochastic_process} (see also Fig.~\ref{fig:interconnection}).
\subsection{Asymptotic Behavior of ECM Dynamics}
We first establish that the ECM concentration $\ecm$ remains bounded and converges to a steady-state value, regardless of the evolution of the cell mass configuration $\Label$. 
We additionally characterize the limiting value of $\ecm$ when the cell mass configuration is held fixed.
Formally, we have the following property.

\begin{lemma}[Asymptotic Behavior of ECM]\label{lem:asymptotic_ecm_behavior}
    If $\left\{\statek\right\}_{\iter \in \natural}$ is a trajectory for the interconnection of~\eqref{eq:ecm_jump_linear}-\eqref{eq:tumor_stochastic_process}, then $\lim_{\iter \to \infty} \ecmk = \ecmlim \in [0, \ecmInit]^{|\grid|}$.
    In particular, if the cell mass configuration is fixed, i.e. $\Labelk = \bar{\Label}$ for all $\iter \in \natural$, then 
    \begin{equation*}
        \lim_{\iter\to\infty} \ecmpk{\gridPoint} =
        \begin{cases}
            0 & \text{if } \A_\gridPoint(\bar{\Label}) \in [0,1)\\[4pt]
            \ecmp{\gridPoint}^0 & \text{if } \A_\gridPoint(\bar{\Label}) = 1
        \end{cases}, \quad \forall \gridPoint \in \grid.
    \end{equation*}
\end{lemma}
\begin{proof}
    See Appendix~\ref{app:proof_lem_ecm_convergence}.
\end{proof}

\subsection{Transition Probabilities of Cell Mass Configuration}\label{subsec:trans_prob_characterization}
 
Let us start by introducing the difference set between two configurations as $\diffset\left(\Label_1, \Label_2\right) \coloneqq \left\{ \gridPoint \in \grid : \Label_{1,\gridPoint} \neq \Label_{2,\gridPoint} \right\}$.
Then, the distance function $\metr{\labelSet}\left(\Label_1, \Label_2\right) = \left|\diffset\left(\Label_1, \Label_2\right)\right|$ counts the number of different points between two configuration.
We state that two distinct configurations $\Label_1$ and $\Label_2$ are adjacent if and only if $\metr{\labelSet}\left(\Label_1, \Label_2\right) = 1$. 
Given the current ECPM state $\state^\iter$, the transition probability $\prob{\Labelkk = \Label^\prime \mid \statek}$ for the system to evolve to the configuration $\Label^\prime$ is zero either if the two configurations are not adjacent, or the two configurations are adjacent with $\diffset\left( \Labelk, \Label^\prime \right) = \{\gridPoint\}$, but they differ by an outlier (an isolated label), i.e. $\Label^\prime_p \neq \Label^\prime_\gridPointNeigh, \; \forall q \in \Neighbour_p$.
Otherwise, the probability of moving to $\Label^\prime$ is always nonzero.
We now characterize these transition probabilities.
To this end, we first introduce some notation. 
For a given configuration $\Label$, let $\numeq(\Label) \leq |\edges|$ denote the number of edges in $\edges$ with the same label, i.e., $\numeq(\Label) \coloneqq |\left\{(\gridPoint, q) \in \edges : \Labelp{\gridPoint} = \Labelp{\gridPointNeigh}\right\}|$.
Next, we partition the set of configurations adjacent to $\Label$ based on the sign of the energy variation. 
Define $\adjConfp(\state)$ as the set of configurations $\Label^\prime$ adjacent to $\Label$ for which the energy change is positive, specifically
\begin{equation*}
    \adjConfp(\state) \coloneqq \left\{\Label^\prime \in \labelSet^{|\grid|} : \diffset\left( \Label, \Label^\prime \right) = \{\gridPoint\}, \,\Delta \E\left(\Label^\prime | \state\right) > 0\right\}.
\end{equation*}
Similarly, define $\adjConfn(\state)$ as the set of configurations $\Label^\prime$ adjacent to $\Label$ for which the energy change is non-positive, i.e.
\begin{equation*}
    \adjConfn(\state) \coloneqq \left\{ \Label^\prime \in \labelSet^{|\grid|} : \diffset\left( \Label, \Label^\prime \right) = \{\gridPoint\}, \,\Delta \E\left(\Label^\prime | \state\right) \leq 0\right\}.
\end{equation*}
Then, all transient configurations adjacent to $\Label$ belong to the set $\adjConf(\state):= \adjConfp(\state) \cup \adjConfn(\state)$.
With these definitions in place, we can now express the transition probabilities as follows.
For a configuration $\Label^\prime \in \adjConfn(\state)$ (favorable energy change), the transition probability is
\begin{equation}\label{eq:trans_prob_favorable}
    \prob{\Labelkk \!= \Label^\prime | \statek} = \frac{1}{|\edges| - \numeq(\Labelk)},
\end{equation}
where the denominator reflects that only adjacent configurations with different labels can be reached in one step, excluding the $\numeq(\Label)$ edges with the same label.
Instead, for a configuration $\Label^\prime \in \adjConfp(\state)$ (unfavorable energy change), the transition probability is
\begin{equation}\label{eq:trans_prob_unfavorable}
    \prob{\Labelkk \!= \Label^\prime | \statek} = \frac{\exp\left(-\Delta \E\left(\statek\right) / \beta\right)}{|\edges| - \numeq(\Labelk)}.
\end{equation}
Finally, the probability of remaining in the same $\Label$ is
\begin{equation}\label{eq:trans_prob_stay}
    \prob{\Labelkk \!= \Labelk \mid \statek} = \displaystyle\sum_{\Label^\prime \in \adjConfp(\statek)}\frac{1 - \exp\left(-\Delta \E\left(\statek\right) / \beta\right)}{|\edges| - \numeq(\Labelk)},
\end{equation}
which accounts for $\Label$-transitions to energetically unfavorable configurations that do not occur.
Based on the above expressions, we can now provide explicit uniform lower and upper bounds for the transition probabilities in terms of the model parameters and grid dimensions.
To state the bound, let $\DeltaHmax, \DeltaHmin \in \real$ be constants such that, for every feasible adjacent configuration $\Label^\prime \in \adjConf(\state)$ it holds that $\DeltaHmin \le \Hamiltonian(\Label^\prime) - \Hamiltonian(\Label) \le \DeltaHmax$.

\begin{lemma}[Bounds on Transition Probabilities]\label{lem:bounds_transition_prob}
    For any trial configuration $\Label^\prime \in \adjConf(\statek)$, the corresponding transition probability admits the uniform lower and upper bounds
    \begin{equation*}
        \probLBound \leq \prob{\Labelkk = \Label^\prime \mid \statek} \leq \frac{1}{2} \quad \text{where} \quad \probLBound := \frac{e^{-(\DeltaHmax + \Kecm \ecmInit)/\beta}}{\xdim \left(8\ydim - 6\right)}.
    \end{equation*}
\end{lemma}
\begin{proof}
    See Appendix~\ref{app:proof_lem_transition_prob_bounds}.
\end{proof}

\subsection{Interconnected System Evolution and Absorbing States}

We finally present the main result of this section, namely the characterization of the interconnected system dynamics formulated in Section~\ref{sec:system_theoretical_viewpoint}. 
This result leverages the properties established in previous lemmas to describe the global system behavior.
Considering the coupled dynamics in~\eqref{eq:ecm_jump_linear}-\eqref{eq:tumor_stochastic_process}, the state space is $\Sigma \times F$, where $\Sigma \subset \labelSet^{|\grid|}$ represents the finite set of label configurations and $F = [0, \ecmInit]^{|\grid|} \subset \real^{|\grid|}_{\geq 0}$ is the compact set of ECM concentrations.
The global system, with state $\statek = (\Labelk, \ecmk)$, satisfies the Markov property. Since $F$ is uncountable, $\left\{\statek\right\}_{\iter \in \natural}$ is a discrete-time \emph{general state-space Markov chain} in the sense of~\cite[Chapter~3]{meyn2012markov}. 
We now characterize the absorbing states and establish convergence properties of the interconnected system.
Specifically, the following theorem extends results from~\cite{voss2012multi} on standard CPMs to the  ECPM framework with ECM dynamics.
\begin{theorem}[Global system convergence]\label{thm:global_convergence}
    The general state-space Markov chain $\left\{\statek\right\}_{\iter \in \natural}$ represented by the ECPM dynamics in~\eqref{eq:ecm_jump_linear}-\eqref{eq:tumor_stochastic_process} converges almost surely to the equilibrium (absorbing) set $\absSet$, given by 
    \begin{equation*}
      \absSet = \absSet_{\labelEmpty} \cup \left(\bigcup_{\labels \in \labelSet \setminus \left\{\labelEmpty\right\}} \absSet_\labels\right)
    \end{equation*}
    in which
    \begin{equation*}
        \absSet_{\labelEmpty} := \left\{\state : \Labelp{\gridPoint} = \labelEmpty, \, \forall \gridPoint \in \grid\right\}, \quad \absSet_\labels := \left\{\state : \Labelp{\gridPoint} = \labels, \ecmp{\gridPoint} = 0, \, \forall \gridPoint \in \grid \right\}.
    \end{equation*}
    Specifically, the $\Label$-dynamics converge with probability 1 in finite time to the uniform configuration set $\uniformConfig$, independently of $\ecm$ and the initial condition $\state^0$, where
    \begin{equation*}
        \uniformConfig := \left\{\Label \in \labelSet^{|\grid|} : \Labelp{\gridPoint} = \Labelp{\gridPointNeigh}, \forall (\gridPoint, q) \in \edges\right\}.
    \end{equation*}
\end{theorem}
\begin{proof}
    See Appendix~\ref{app:proof_thm_global_convergence}.
\end{proof}

\section{Invasion Time in One-Dimensional ECPM} \label{sec:1Dgrid_absorption_times}

\begin{figure}[htpb!]
    \centering
    \includegraphics[width=0.65\columnwidth]{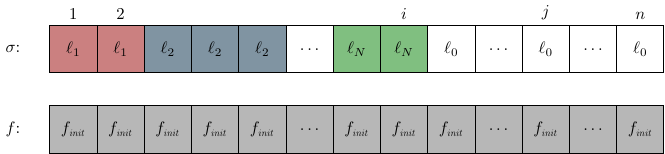}
    \caption{%
        Illustration of the initial ECPM state for a 1D grid with
        $\numCells$ tumor cells surrounded by ECM.
        The indices $\indexStart$, $\indexStop > \indexStart$, and $\latticeDim$ denote
        the initial front depth, target depth, and grid length, respectively.
    }
    \label{fig:init_state_1Dgrid_cells}
\end{figure}

In this section, we analyze the ECPM in a one-dimensional (1D) lattice, focusing on the \emph{invasion time}, i.e. the time steps required for the tumor front to reach a specific lattice site given an initial configuration. 
As highlighted in~\cite{yamada2019extracellular},
the 1D-ECPM can be viewed as a reduced representation of tumor invasion, capturing migration dynamics that closely mirror the radial progression observed in three-dimensional settings.
Moreover, we restrict admissible $\Label$-updates to forward invasion, neglecting reverse copy attempts. 
This assumption is motivated as follows.
Biologically, 1D migration along narrow ECM fibers exhibits stable polarity with predominant forward motion driven by matrix degradation at the leading edge~\cite{yamada2019extracellular}. 
Analytically, this restriction provides a conservative estimate: by preventing empty sites from overwriting tumor cells, we obtain a lower bound on the actual mean invasion time.
Let $\grid$ be such that $\xdim = 1$ and $\ydim = \latticeDim$, with $\latticeDim$ being the number of lattice sites.
Let the initial ECM density be uniform across the lattice, i.e.
\begin{equation}\label{eq:init_cond_ecm}
    \ecmp{\gridPoint}^0 = \ecmInit, \quad \forall \gridPoint \in \grid,
\end{equation}
where $\ecmInit \in \real_{> 0}$ denotes the initial ECM density at a lattice site.
We denote by $\ecm^0$ the initial ECM vector satisfying~\eqref{eq:init_cond_ecm}.
It follows that $\ecm^k_p \in [0, \ecmInit]$ for all $k \in \natural$ and $p \in \grid$ (see Lemma~\ref{lem:asymptotic_ecm_behavior}). 
Now, we define the invasion time as follows.
\begin{definition}[Invasion time]\label{def:absTime}
    Given a grid point $\indexStop \in \grid$, we define the maximum tumor depth at iteration $\iter$ as $\maxTumorDepth{\Labelk} := \max \left\{y_\gridPoint \in \Yset : \Labelp{\gridPoint}^\iter \neq \labelEmpty\right\}$, where $y_\gridPoint$ denotes the vertical coordinate of $\gridPoint \in \grid$.
    The invasion time $\absTime{\indexStop}$ is defined as the minimum number of ECPM update steps required for the tumor front to reach the lattice site at depth $\indexStop$, i.e.
    \begin{equation*}\label{eq:inv_time}
        \absTime{\indexStop} := \min  \left\{\iter \geq 0 : \maxTumorDepth{\Labelk} \geq \indexStop\right\}.
    \end{equation*}
\end{definition}
We now introduce a compact formulation of the invasion time probability.
Let $\probAbsTime{\indexStop}{\indexStart}{\iter}{\state^0}$ indicate the probability that the tumor reaches lattice site $\indexStop$ in $\iter$ steps, starting from initial ECPM state $\state^0$ with maximum tumor depth $\indexStart \leq \indexStop$. 
Formally,
\begin{equation}\label{eq:absTime_prob}
\probAbsTime{\indexStop}{\indexStart}{\iter}{\state^0}
:= \prob{\absTime{\indexStop} = \iter \mid \maxTumorDepth{\LabelInit} = \indexStart, \ecm^0}.
\end{equation}
Building on Definition~\ref{def:absTime}, the mean invasion time is defined as the conditional expectation of $\absTime{\indexStop}$ given the initial ECPM state. Since $\absTime{\indexStop}$ is a discrete random variable taking values $\iter \in \integer_{>0}$ with probability mass function~\eqref{eq:absTime_prob}, we obtain
\begin{equation}\label{eq:mean_abs_time}
    \meanAbsTime{\indexStop}{\indexStart}\!\left(\state^0\right)
    := \expect{\absTime{\indexStop}
    \mid \maxTumorDepth{\LabelInit} = \indexStart,\; \ecm^0} 
    = \sum_{\iter=1}^{\infty} \iter \,
    \probAbsTime{\indexStop}{\indexStart}{\iter}{\state^0},
\end{equation}
provided that the series converges. 
In practice, the mean invasion time can be approximated by truncating the infinite series in~\eqref{eq:mean_abs_time} to a finite horizon $\truncHorz \in \natural$, yielding
\begin{equation}\label{eq:mean_abs_time_truncated}
    \meanAbsTime{\indexStop}{\indexStart}\left(\state^0\right)
    \approx \sum_{\iter=1}^{\truncHorz} \iter \; \probAbsTime{\indexStop}{\indexStart}{\iter}{\state^0}.
\end{equation}
We now proceed with the derivation of closed-form expressions for the invasion time probability $\probAbsTime{\indexStop}{\indexStart}{\iter}{\state^0}$ for $\numCells \geq 1$ tumor cells.
Let us consider a 1D-ECPM initialized in state $\state^0$, as depicted in Figure~\ref{fig:init_state_1Dgrid_cells}. In detail, the initial configuration consists of $\numCells$ contiguous tumor cells, labeled $\labels_1, \ldots, \labels_\numCells$, occupying the lattice $\grid$. The tumor front is initially at position $\maxTumorDepth{\LabelInit} = \indexStart$, and each cell $\labels$ has an initial volume $\Vol^0_\labels$.
In this framework, the label and type sets are $\labelSet = \left\{\labelEmpty,\, \labels_1,\, \ldots,\, \labels_\numCells\right\}$ and $\typeSet = \left\{\typeECM,\, \typeTumor\right\}$, where $\labelEmpty$ corresponds to empty lattice sites covered by ECM,
$\typeECM$ and $\typeTumor$ denote the ECM and tumor cell types respectively, with $\Type_{\labelEmpty} = \typeECM$ and
$\Typel{\labels} = \typeTumor$ for all $\labels \neq \labelEmpty$.
Moreover, since only tumor cells secrete degrading enzymes (i.e.\
$\kecm_{\typeECM} = 0$), the dynamics in~\eqref{eq:ecm} simplify to
\begin{equation}\label{eq:ecm_1D}
    \ecmpkk{\gridPoint}
    = \Bigl(1 - \kecm_{\typeTumor}\,
    \enzime\!\left(\gridPoint, \typeTumor, \Labelk\right)\Bigr)
    \ecmpk{\gridPoint},
    \quad \forall \gridPoint \in \grid,
\end{equation}
with the proteolytic enzyme secreted at site $\gridPoint$ (see general expression in~\eqref{eq:enzime}), given by
\begin{equation*}
    \enzime\!\left(\gridPoint, \typeTumor, \Label\right)
    = \kenzime{\typeTumor}
    \sum_{\d=1}^{\dmax_{\typeTumor}}
    \nearcells\!\left(\gridPoint, \typeTumor, \Label\right)
    e^{-\enzimeDecay{\typeTumor}\,\d}.
\end{equation*}
Before presenting the main result, we introduce some definitions needed to express the invasion time probability in a compact form.
We denote by
\begin{equation}\label{eq:binom_compact_def}
    b(\iter,\indexStop,\indexStart) := \binom{\iter-1}{\indexStop-\indexStart-1},
\end{equation}
the number of ordered partitions of $\iter$ steps into $\indexStop-\indexStart$ forward transitions of the tumor front, namely transitions such that $\maxTumorDepth{\Labelk} = \indexStart + \transIter$ for some $\transIter \in \{1,\ldots,\indexStop-\indexStart\}$.
Each partition is indexed by $\partIter \in \{1,\ldots,b(\iter,\indexStop,\indexStart)\}$ and denoted by $(\iter_1^{\partIter},\ldots,\iter_{\indexStop-\indexStart}^{\partIter})$, such that
\begin{equation}\label{eq:partition_k}
    \iter = \sum_{\transIter=1}^{\indexStop-\indexStart} \iter_\transIter^{\partIter},
    \quad \iter_\transIter^{\partIter} \ge 1,
\end{equation}
in which the index $\transIter$ labels the successive front transitions.
Additionally, within each partition $\partIter$ we define the cumulative iteration counts $\cumulIter := \sum_{g=1}^{\transIter} \iter_{g}^{\partIter}$,
so that the $\transIter^{\text{th}}$ transition in the $\partIter^{\text{th}}$ partition occurs between iterations $\zeta_{\transIter-1}^{\partIter}$ and $\cumulIter - 1$.
The following theorem characterizes the invasion time for the setting introduced above.

\begin{theorem}[Invasion Time Probability for $\numCells$ Tumor Cells in a 1D-ECPM]\label{theorem:absTime_prob_multicell}
    Consider a 1D lattice with an initial configuration $\LabelInit$ consisting of $\numCells \geq 1$ contiguous tumor cells, where the tumor front is at starting position $\maxTumorDepth{\LabelInit} = \indexStart$ and let $\indexStop > \indexStart$ be a target front position.
    Then, the probability that the tumor front reaches site $\indexStop$ from $\indexStart$ in exactly $\iter \geq \indexStop - \indexStart$ steps is given by
    \begin{equation}\label{eq:absTime_prob_from_i_to_j_in_k_steps}
        \probAbsTime{\indexStop}{\indexStart}{\iter}{\state^0}
        = \sum_{\partIter=1}^{b(\iter,\indexStop,\indexStart)}
        \prod_{\transIter=\indexStart}^{\indexStop-1}
        \underbrace{\probAbsTime{\transIter+1}{\transIter}{\iter_{\transIter - \indexStart + 1}^{\partIter}}{\state^{\transIter - \indexStart}}}_{=:\wp_\transIter^\partIter}.
    \end{equation}
    Here, each factor $\wp_\transIter^\partIter$ corresponds to the probability of the $(\transIter - \indexStart + 1)^{\text{th}}$ transition within the $\partIter^{\text{th}}$ partition to occur, i.e.
    \begin{equation}\label{eq:absTime_prob_ks_multicell}
        \wp_\transIter^\partIter = \frac{\prob{ \cdot \mid \state^{\cumulIter-1}}}{2\numCells-1}
        \prod_{\hIter=\zeta_{\transIter-1}^{\partIter}}^{\cumulIter-1} \!
        \left(1 - \frac{\prob{\cdot \mid \state^{\hIter}}}{2\numCells-1}\right),
    \end{equation}
    where, at each intermediate step $\hIter$, $\prob{\cdot \mid \state^{\hIter}}$ is the transition probability in~\eqref{eq:transition_prob}, with energy variation defined by
    \begin{equation}\label{eq:deltaE_h_multiplecells}
        \Delta\E\left(\state^\hIter\right)
        = \lambda\!\left[1 + 2\!\left(\Vol^{\hIter}_{\labels_\numCells} - \targetVol\right)\right]
        + \Kecm\!\left(\ecmp{\gridPoint}^{\hIter} - \ecmp{\gridPointNeigh}^{\hIter}\right),
    \end{equation}
    where $\gridPoint$ and $\gridPointNeigh$ denote, respectively, the current front position $\maxTumorDepth{\Label^\hIter}$ and its adjacent empty-site neighbor.
\end{theorem}

\begin{proof}
    See Appendix~\ref{app:proof_thm_absTime_prob_multicell}.
\end{proof}

\section{Numerical Simulations and Validation}\label{sec:simulations}

In this section, we validate the analytical results from Section~\ref{sec:1Dgrid_absorption_times} via numerical experiments. 
We begin by detailing the simulation setup and parameter choices. 
Next, we conduct a sensitivity analysis of the mean invasion time, comparing empirical results against closed-form solutions, across key ECPM parameters and various invasion distances. 
\subsection{Simulation Details}

All simulations were implemented in \textsc{Python} programming language by modeling the coupled tumor-ECM dynamics described in Section~\ref{sec:system_theoretical_viewpoint}. 
Each run starts from an initial configuration $\state^0$, corresponding to the framework illustrated in Figure~\ref{fig:init_state_1Dgrid_cells}. 
To choose $\state^0$, we resorted to typical experimental setups in which tumor spheroids are seeded in a collagen matrix.
Specifically, recalling that 1D scenarios mimic the cross-section of a 3D tumor spheroid, we considered an initial state in which a contiguous block of cells $\numCells$ occupies $50\%$ of the 1D-lattice, for instance $\numCells = 3$ cells with initial volumes $\initVol=10$ in a lattice of length $\latticeDim = 60$.

We now specify the ECPM parameter values used in the simulations.
The environmental noise parameter in~\eqref{eq:transition_prob}, weighing the probability of energetically unfavorable events, was set to $\beta = 4$. 
We set the target cell volume to $\targetVol = 2\initVol$ (see~\eqref{subeq:HamV}) to allow for significant morphological expansion during invasion. This choice provides the cell with sufficient volumetric freedom to deform, elongate, and invade the surrounding medium without being overly constrained by the volume conservation term in the Hamiltonian.
Adhesion energies were chosen as in~\eqref{eq:interaction_energy_standard}, specifically $\Jcc = 3$ and $\Jec = 6$ considering a stronger adhesion between tumor cells and ECM than between tumor cells themselves (typical of invasive phenotypes). However, since the invasion time probability in Theorem~\ref{theorem:absTime_prob_multicell} does not depend on adhesion energies, these values do not affect the analytical results.
The enzyme decay rate was set to $\enzimeDecay{\typeTumor} = 2$, while the product between the secretion and ECM degradation rates in~\eqref{eq:ecm} was set to $\kenzime{\typeTumor} \cdot \kecm_{\typeTumor} = 0.01$, and the maximal influence radius to $\dmax_{\typeTumor} = 4$ (see~\eqref{eq:enzime}).
The adopted ECPM parameter set is consistent with standard ECPM configurations reported in the literature~\cite{graner1992simulation, turner2002intercellular, turner2004tamoxifen}, suitably adapted to the one-dimensional framework for validation purposes.

To estimate the mean invasion time $\meanAbsTime{\indexStop}{\indexStart}$ we performed $10\,000$ Monte Carlo simulations of the ECPM dynamics in~\eqref{eq:ecm_jump_linear}-\eqref{eq:tumor_stochastic_process}, adopting the stopping criterion $\maxTumorDepth{\Labelk} = \indexStop$. 
Here, $\indexStop$ denotes the target lattice site reached by the tumor front, while $\indexStart$ represents its initial position (i.e., $\maxTumorDepth{\LabelInit} = \indexStart$). Each simulation terminates when the front reaches site $\indexStop > \indexStart$, and the corresponding invasion time $\absTime{\indexStop}$ is recorded.
Moreover, to control computational cost, we set a maximum number of iteration steps $\truncHorz \in \natural$. 
This horizon is chosen to capture most of the probability mass of the invasion-time distribution, ensuring accurate estimates of the mean invasion time.
Details on the stopping criterion used to select $\truncHorz$ are provided in the next subsection.

\subsection{Sensitivity of Invasion Time to $\lambda$ and $\Kecm$}\label{subsec:mean_time_distance}

We now investigate how the mean invasion time $\meanAbsTime{\indexStop}{\indexStart}\!\left(\state^0\right)$ scales with the invasion distance $\indexStop - \indexStart$, and how this scaling is shaped by the two key parameters entering the energy term~\eqref{eq:deltaE_h_multiplecells}, namely the volume-constraint weight $\lambda$ and the haptotactic coupling gain $\Kecm$. 
To validate the closed-form expression established in Theorem~\ref{theorem:absTime_prob_multicell} against empirical methods, for every parameter configuration we compare three \emph{mean invasion time estimators}:
\begin{itemize}
    \item \emph{Empirical (full model)}: sample mean over Monte Carlo runs of the ECPM dynamics defined in Section~\ref{sec:system_theoretical_viewpoint}, in which ECM-over-tumor interactions are bidirectional, and the tumor front can advance or retreat at each step;
    \item[\scriptsize$\blacksquare$] \emph{Empirical (simplified model)}: sample mean over Monte Carlo runs of the forward-only restricted dynamics introduced in the analysis of Section~\ref{sec:1Dgrid_absorption_times};
    \item[\scriptsize$\blacktriangle$] \emph{Closed-form (simplified model)}: the truncated analytical mean of~\eqref{eq:mean_abs_time_truncated}, evaluated via Theorem~\ref{theorem:absTime_prob_multicell}.
\end{itemize}

\begin{figure}[htpb!]
    \centering
    \includegraphics[width=0.6\columnwidth]{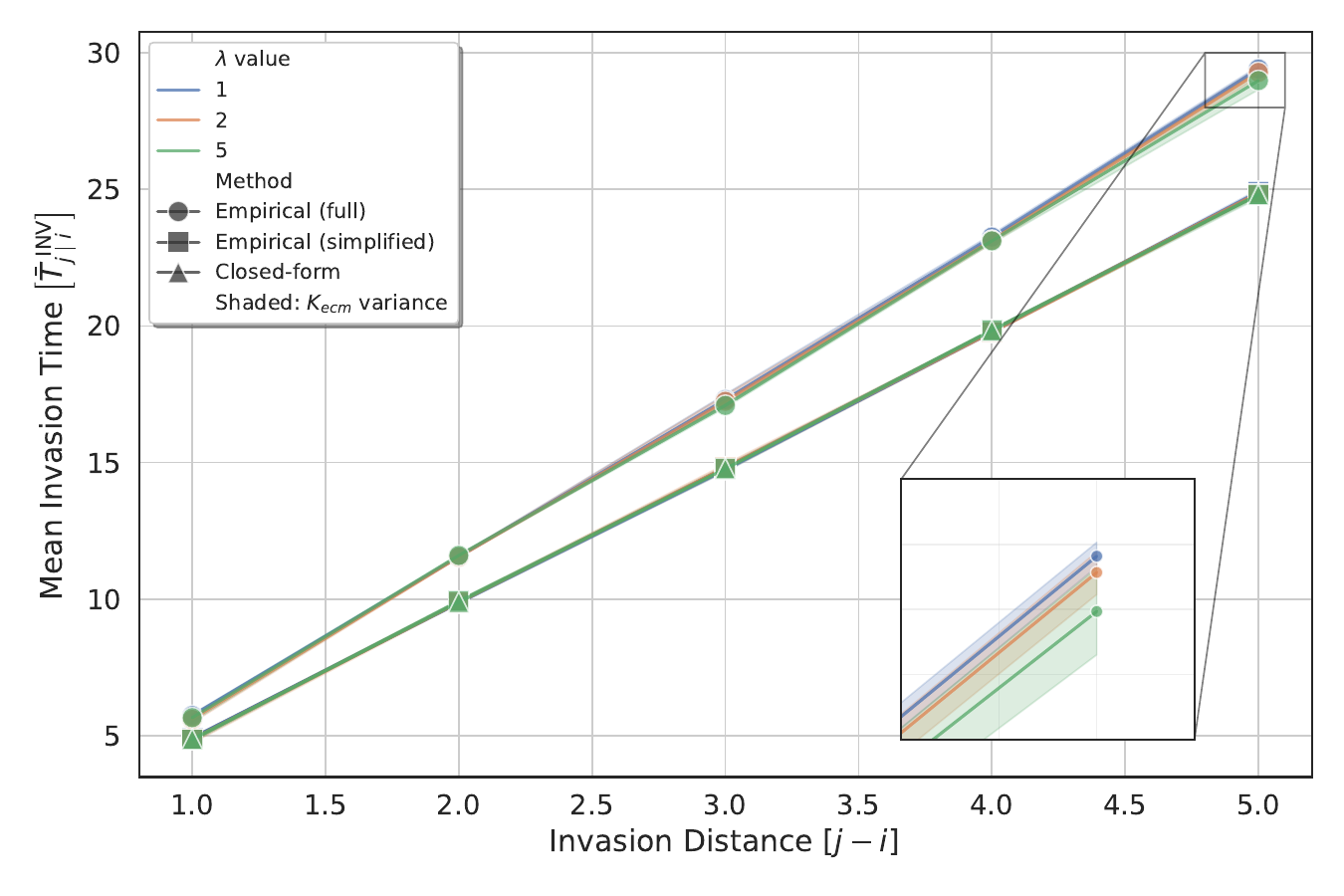}
    \caption{Mean invasion time estimations versus invasion distance $\indexStop - \indexStart$ for several values of $\lambda$. Shaded areas denote the variance induced by varying $\Kecm$ around each $\lambda$ curve. 
    }
    \label{fig:mean_time_multi_lambda_kecm_variance}
\end{figure}

\begin{figure}[htpb!]
    \centering
    \includegraphics[width=0.6\columnwidth]{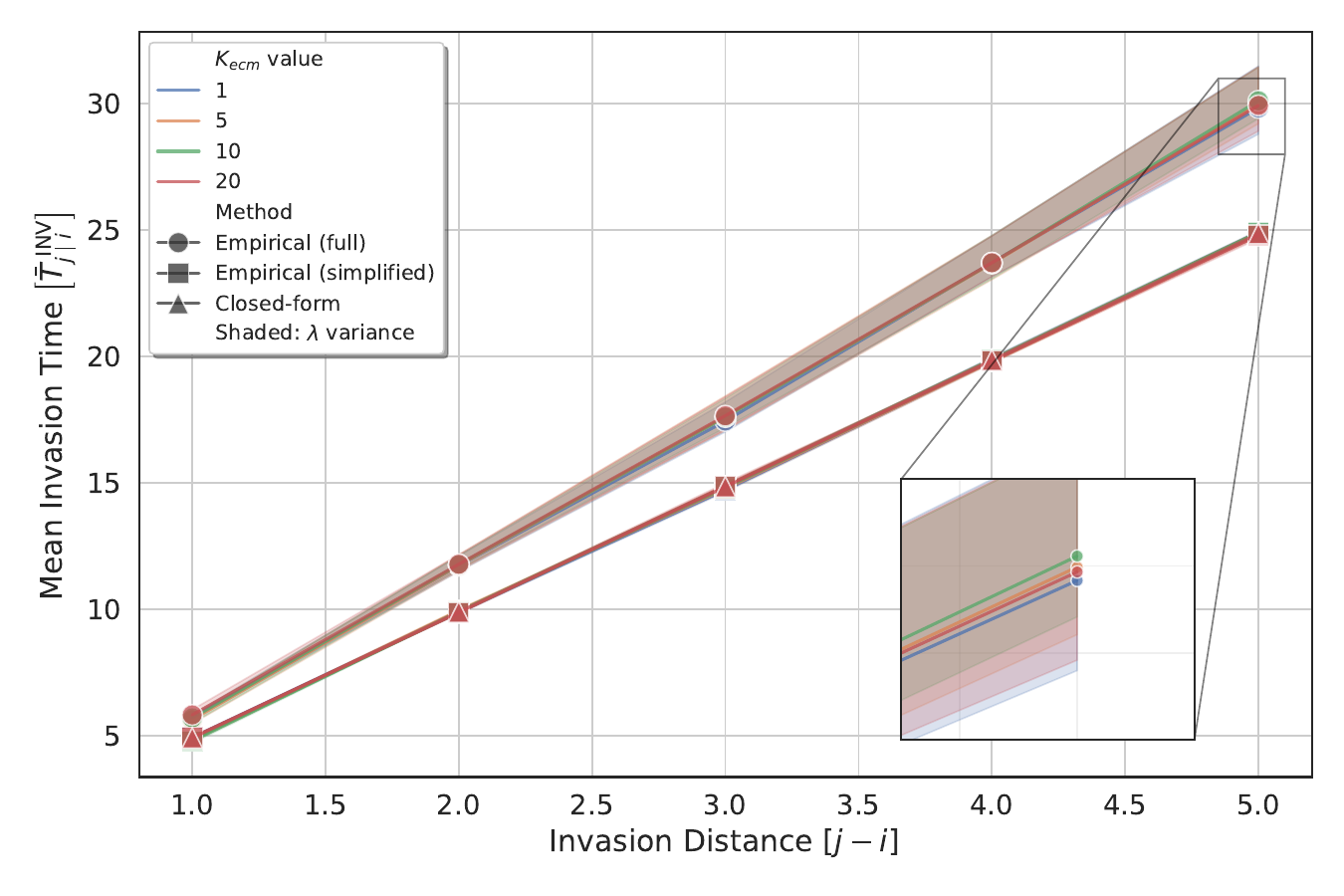}
    \caption{Mean invasion time estimations versus invasion distance $\indexStop - \indexStart$ for several values of $\Kecm$. Shaded areas denote the variance induced by varying $\lambda$ around each $\Kecm$ curve. 
    }
    \label{fig:mean_time_multi_kecm_lambda_variance}
\end{figure}

\begin{figure*}[htpb!]
    \centering
    \includegraphics[width=\textwidth]{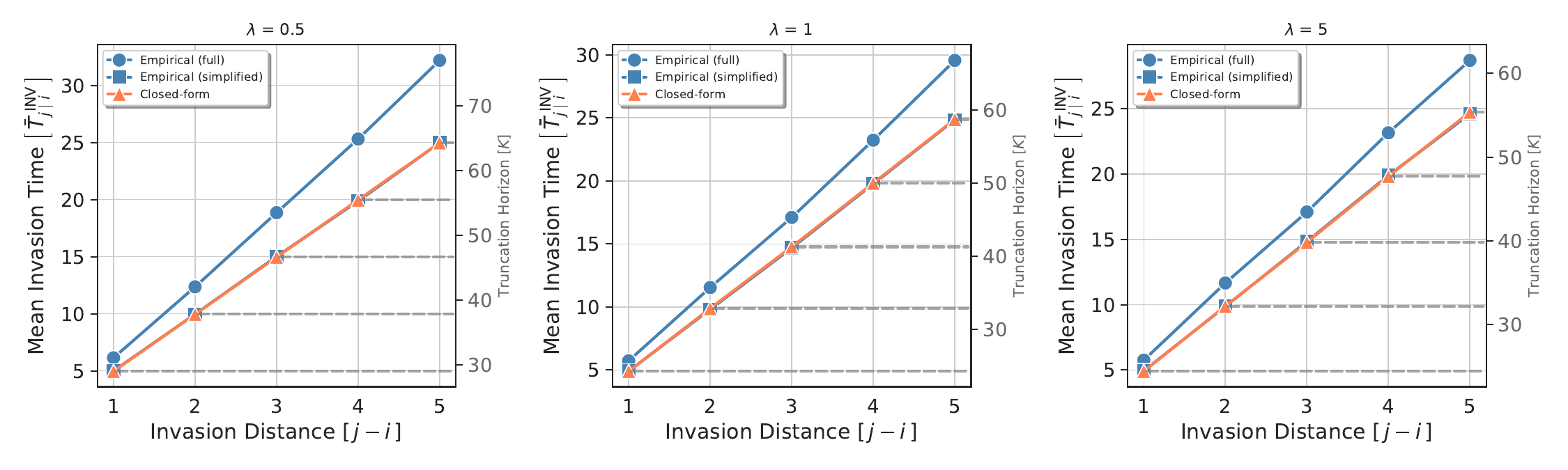}
    \caption{Mean invasion time versus invasion distance $\indexStop - \indexStart$ for different $\lambda$ values, comparing the empirical (full model), empirical (simplified model), and closed-form estimators. The right axis (gray dashed line) reports the truncation horizon $\truncHorz$ satisfying~\eqref{eq:adaptive_truncation_rule} for each invasion distance.}
    \label{fig:mean_time_lambda_subplots}
\end{figure*}

For each parameter configuration, the truncation horizon $\truncHorz$ used to evaluate~\eqref{eq:mean_abs_time_truncated} is selected via a practical heuristic. 
We therefore choose $\truncHorz$ as the smallest integer satisfying cumulative-probability and point-mass thresholds for both the empirical and closed-form distributions, i.e.
\begin{equation}\label{eq:adaptive_truncation_rule}
    \sum_{\iter=1}^{\truncHorz} \probAbsTime{\indexStop}{\indexStart}{\iter}{\state^0} \geq 0.98, \quad
    \probAbsTime{\indexStop}{\indexStart}{\truncHorz}{\state^0} < 10^{-5},
\end{equation}
in which the second condition is checked only after a sufficiently large $\truncHorz$ ensuring that the majority of the probability mass is captured.
This guarantees a comparable level of distributional coverage across all the parameter combinations considered below, even though the resulting $\truncHorz$ value itself varies substantially with $\lambda$, $\Kecm$, and the invasion distance, as shown next.
We analyze the relative influence of $\lambda$ and $\Kecm$ on the mean invasion time. 
Figure~\ref{fig:mean_time_multi_lambda_kecm_variance} reports $\meanAbsTime{\indexStop}{\indexStart}$ as a function of the invasion distance for several values of $\lambda$; the shaded band around each curve shows the family of functions obtained by varying $\Kecm\in[1,20]$. 
Figure~\ref{fig:mean_time_multi_kecm_lambda_variance} provides the complementary view, grouping curves by $\Kecm$ and showing shaded bands for $\lambda\in[1,5]$. 
The closed-form estimator is independent of both $\lambda$ and $\Kecm$, and it coincides with the empirical estimator obtained from the simplified model, thereby validating our analytical derivation. 
Notably, estimation quality decreases with increasing invasion distance, which is consistent with a higher incidence of tumor retreat events, that explains the growing discrepancy between estimators for large invasion distances.
To assess the accuracy of the closed-form expression from another viewpoint, Figure~\ref{fig:mean_time_lambda_subplots} reports the three estimators introduced above, for three representative values $\lambda \in \{0.5, 1, 5\}$, together with the corresponding truncation horizon $\truncHorz$ satisfying conditions~\eqref{eq:adaptive_truncation_rule}.
\section{Conclusions}\label{conclusions}

In this paper, we introduced a system-theoretic framework for analyzing an ECPM for tumor invasion by reformulating it as an interconnected system that couples a stochastic process describing locally interacting tumor cells with a jump-linear system modeling the ECM environment.
Within this formulation, we characterized the asymptotic behavior of both subsystems, derived transition probabilities and uniform bounds, and exploited the sparse structure of the energy variation terms.
These results yielded global convergence guarantees, showing that the coupled dynamics reach equilibrium states independently of the ECM field.
In one-dimensional settings, we derived a closed-form for invasion-time probabilities that depend on local ECM gradients, number of cells and target cell volume.
Numerical simulations validated the theory across multiple scenarios and supported the sensitivity analysis of the mean invasion time with respect to key ECPM parameters and invasion distances, providing an analytical foundation for tumor invasion dynamics in ECM substrates.

\appendix
\subsection{Proof of Lemma~\ref{lem:energy_change_decomposition}}\label{app:proof_lem_energy_decomposition}

The adhesion-energy change $\Delta \HamJ = \HamJ(\Label^\prime) - \HamJ(\Label)$ is computed considering all pairwise interactions between neighboring sites (see~\eqref{subeq:HamJ}).
However, since only the interaction terms involving site $q$ and thus $\gridPoint \in \Neighbour_{q}$ change, all other terms cancel out, leading to equation~\eqref{eq:deltaHJ_simplified}.
For the volume-energy change $\Delta \HamV = \HamV(\Label^\prime) - \HamV(\Label)$, after the copy attempt from $\gridPoint$ to $q$, the affected cell volumes evolve according to $\Vol_\labels^+ = \Vol_\labels + \rho(\Labelp{\gridPoint},\labels)$, $ \forall \labels \in \{\Labelp{\gridPoint},\Labelp{\gridPointNeigh}\}$,where $\rho(\labels_1,\labels_2) = 2\KroneckerDelta(\labels_1,\labels_2)-1$ accounts for the increase or decrease in volume. 
All the other cell volumes remain unchanged, thus canceling out in the energy difference.
Therefore, the Hamiltonian volume energies at $\Label$ and $\Label^\prime$ can be simplified as
\begin{align*}
    \HamV(\Label) &= \lambda \sum_{\labels \in \{\Labelp{\gridPoint},\Labelp{\gridPointNeigh}\}} (\Vol_\labels - \targetVol)^2, \\
    \HamV(\Label^\prime) &= \lambda \sum_{\labels \in \{\Labelp{\gridPoint},\Labelp{\gridPointNeigh}\}} (\Vol_\labels + \rho(\Labelp{\gridPoint},\labels) - \targetVol)^2.
\end{align*}
Then, subtracting these two expressions, developing the squares, and simplifying leads to the following volume-energy change
\begin{equation*}
    \Delta \HamV(\Label) = \lambda \sum_{\labels \in \{\Labelp{\gridPoint},\Labelp{\gridPointNeigh}\}} \left[\rho(\Labelp{\gridPoint},\labels)^2 + 2\rho(\Labelp{\gridPoint},\labels)\left(\Vol_\labels - \targetVol\right)\right]
    = \lambda \sum_{\labels \in \{\Labelp{\gridPoint},\Labelp{\gridPointNeigh}\}} \left[1 + 2\rho(\Labelp{\gridPoint},\labels) (\Vol_\labels - \targetVol)\right],
\end{equation*}
which corresponds to equation~\eqref{eq:deltaHV_simplified} where the empty label $\labelEmpty$ is excluded through the term $(1-\KroneckerDelta(\labelEmpty,\labels))$.
The ECM-dependent term $\Kecm\left(\ecmp{\gridPoint} - \ecmp{\gridPointNeigh}\right)$ arises directly from definition~\ref{step:DeltaE} in Algorithm~\ref{alg:MMA}, thus completing the decomposition of $\Delta \E(\Label, \ecm)$ stated in the Lemma.

\subsection{Proof of Lemma~\ref{lem:asymptotic_ecm_behavior}}\label{app:proof_lem_ecm_convergence}

Considering the dynamics of $\ecmk$, we have for all $\gridPoint \in \grid$ that
\begin{equation*}
    0 \leq \ecmpkk{\gridPoint} =
    \underbrace{\A_\gridPoint(\Labelk)}_{\in [0, 1]} \ecmpk{\gridPoint} 
    \leq \ecmpk{\gridPoint} 
    \leq \ecmp{\gridPoint}^0.
\end{equation*}
Hence, by the monotone convergence theorem the sequence admits a limit in the interval $[0, \ecm_p^0], \; \forall \gridPoint \in \grid$.
When $\Labelk = \bar{\Label}$ for all $\iter \in \natural$, the diagonal structure of $\A(\bar{\Label})$ yields $\ecmpk{\gridPoint} = \left[\A_\gridPoint(\bar{\Label})\right]^\iter \ecmp{\gridPoint}^0$.
The stated limit follows since $\A_\gridPoint(\bar{\Label}) \in [0,1]$ determines whether the sequence decays to zero or remains constant.

\subsection{Proof of Lemma~\ref{lem:bounds_transition_prob}}\label{app:proof_lem_transition_prob_bounds}

Consider a trial cell mass configuration $\Label^\prime$ such that $\Label^\prime_\gridPointNeigh = \Labelp{\gridPoint}$, with $\gridPoint \in \grid$ and $q \in \Neighbour_{\gridPoint}$.  
From the proof of Lemma~\ref{lem:asymptotic_ecm_behavior}, it follows that the ECM variation satisfies $\left(\ecmk_{\gridPoint} - \ecmk_\gridPointNeigh\right) \in [-\ecmInit,\, \ecmInit] \; \forall\, \iter \geq 1$,
where $\ecmInit$ is the initial uniform ECM concentration.
By definition of $\DeltaHmax$ and $\DeltaHmin$, the total energy variation can be bounded as follows
\begin{equation*}
    \Delta \E \in
    \left[\DeltaHmin - \Kecm \ecmInit,\,
        \DeltaHmax + \Kecm \ecmInit \right].
\end{equation*}
To establish the denominator, we calculate the total number of ordered neighboring pairs in $\edges$. Each interior point $(x, y)$, namely $x \in \Xset$ and $y \in \{2, \ldots, \ydim-1\}$, has $8$ nearest neighbors. 
Instead, the top/bottom boundary points at $y = 1$ and $y = \ydim$ each have $\xdim$ points with $5$ nearest neighbors.  
It follows that $|\edges| = \xdim(\ydim - 2) \cdot 8 + 2\xdim \cdot 5 = \xdim(8\ydim - 6)$ which reduces to $|\edges| = \xdim(\ydim - 2) \cdot 5 + 2\xdim \cdot 3 = \xdim(5\ydim - 4)$ and $|\edges| = \xdim(\ydim - 2) \cdot 2 + 2\xdim \cdot 1 = \xdim(2\ydim - 2)$ for $\xdim = 2$ and $\xdim = 1$, respectively.
These scenarios can be included by defining $m := \min(\xdim, 3)$
to then obtain $|\edges| = \xdim \left[(\ydim-2)(3m-1) + 2(2m-1)\right]$.
Note that, in the Lemma statement, we use the upper bound $|\edges| \leq \xdim(8\ydim - 6)$ for simplicity.
From expressions~\eqref{eq:trans_prob_favorable} and~\eqref{eq:trans_prob_unfavorable}, a $\Label$-dependent lower bound on the transition probability would involve the denominator $|\edges| - \numeq(\Label)$, since only edges with differing labels are feasible.
To obtain an absolute lower bound we consider the worst case $\numeq(\Label) = 0$, i.e., all neighboring pairs carry distinct labels, yielding the largest possible denominator $|\edges| = \xdim(8\ydim - 6)$.
Substituting the worst-case energy increase and denominator in~\eqref{eq:trans_prob_unfavorable} directly yields the stated lower bound for any $\Label^\prime \in \adjConfp(\Label, \ecm)$. The same bound applies to any $\Label^\prime \in \adjConfn(\Label, \ecm)$ since in that case the transition probability~\eqref{eq:trans_prob_favorable} is higher.
For the upper bound, observe that $\numeq(\Label) \leq |\edges| - 2$ always holds, since at least two edges with different labels exist unless the configuration is absorbing. Thus, the denominator is at least $2$, yielding the uniform upper bound $1/2$ for the transition probability. 
Precisely, the upper bound is $1/\varphi$ with $\varphi \geq 2$ depending on the specific boundary conditions and metric defined on $\grid$.

\subsection{Proof of Theorem~\ref{thm:global_convergence}}\label{app:proof_thm_global_convergence}

We begin by identifying the equilibria of the ECPM dynamics in~\eqref{eq:ecm_jump_linear}-\eqref{eq:tumor_stochastic_process}, which correspond to the absorbing states of the Markov chain $\left\{\statek\right\}_{\iter \in \natural}$.

\textit{Case 1 ($\Labelp{\gridPoint} = \labelEmpty, \; \forall \gridPoint \in \grid$) :}
When all lattice sites carry the empty label $\labelEmpty$, no tumor cell occupies any site. By definition of the enzyme model~\eqref{eq:enzime}, cells of type $\typeECM$ produce no degrading enzyme (i.e., $\kecm_{\typeECM} = 0$), so $\A_\gridPoint(\Label) = 1$  for all $\gridPoint \in \grid$. Consequently, the ECM field remains constant at its current value $\ecmp{\gridPoint} \in [0, \ecmp{\gridPoint}^0]$, yielding the absorbing subset $\absSet_{\labelEmpty}$.

\textit{Case 2 ($\Labelp{\gridPoint} = \labels \neq \labelEmpty, \; \forall \gridPoint \in \grid$):}
When every lattice site is occupied by a tumor cell of label $\labels$, the degradation factor satisfies $\A_\gridPoint(\Label) < 1$ for all $\gridPoint \in \grid$, since each site lies within the influence radius of at least one tumor cell, implying $\alpha_\gridPoint(\Label) > 0$ in~\eqref{eq:trans_matrix_local}. 
Therefore, by the fixed-configuration case in Lemma~\ref{lem:asymptotic_ecm_behavior}, the ECM concentration at every lattice point decays monotonically to zero. Once $\ecm = 0_{|\grid|}$, the ECM field is frozen and the configuration $\left(\Label, 0_{|\grid|}\right)$ is absorbing, yielding the subset $\absSet_\labels$.

We now show that the system trajectories converge almost surely to the absorbing set $\absSet$.
By Lemma~\ref{lem:asymptotic_ecm_behavior}, $\ecmk$ converges regardless of $\Label$, allowing us to treat the $\Label$-dynamics as a time-varying process whose transition probabilities evolve through $\ecmk$.
Therefore, the cumulative transition matrix of the $\Label$-dynamics up to step $\iter$ admits the canonical block form   
\begin{equation*}
    \begin{pmatrix} 
        \displaystyle\prod_{s=0}^{k} Q_s & \tilde{R}_k \\ 0 & I 
    \end{pmatrix},
    \quad \tilde{R}_k := \sum_{h=0}^{k}\left(\prod_{s=0}^{h-1} Q_s\right) R_h,
\end{equation*}
where $Q_s$ is the sub-stochastic transition matrix among transient states at step $s$,
$R_h$ is the transition matrix from transient to absorbing states at step $h$, and $I$
is the identity on absorbing states.
The sum of the elements of the $i$-th row of the matrix 
\begin{equation*}
    B(t, m) \coloneqq \prod_{k = t}^{t + m - 1} Q_k,
\end{equation*}
represents the probability of not being absorbed for $m$ steps starting from transient state $i$ at time $t$. Since, by Lemma~\ref{lem:bounds_transition_prob} each  1-step non-zero transition probability is lower bounded by the positive constant $\probLBound$, and that, for each transient state $i$, there exists an admissible path of length $m_i$ to an absorbing state, if we set $M \coloneqq \max_i\{m_i\}$, we have
\begin{equation}\label{eq:abs_prob_bound}
    \infnorm{B(t, M)} = \max_i \sum_j [B(t, M)]_{ij} \leq 1 - \probLBound^M.
\end{equation}
Then, since the iteration step can be decomposed as $\iter = M \lfloor \iter/M \rfloor + \iter \mod M$, we can write
\begin{align*}
    \infnorm{\prod_{s = 0}^{\iter} Q_s} &\stackrel{(a)}{\leq} \left(\prod_{s = 0}^{\lfloor \iter/M \rfloor - 1} \infnorm{\prod_{t = 0}^{M - 1} Q_{Ms + t}}\right) \!\prod_{s = \lfloor \iter/M \rfloor M}^\iter \!\infnorm{Q_s}
    = \left(\prod_{s = 0}^{\lfloor \iter/M \rfloor - 1} \infnorm{B(Ms, M)}\right) \!\prod_{s = \lfloor \iter/M \rfloor M}^\iter \!\infnorm{Q_s}\\
    &\stackrel{(b)}{\leq} \left(1 - \probLBound^M\right)^{\lfloor \iter/M \rfloor} \prod_{s = \lfloor \iter/M \rfloor M}^\iter\infnorm{Q_s}
    \stackrel{(c)}{\leq} \left(1 - \probLBound^M\right)^{\lfloor \iter/M \rfloor} \left(\frac{n}{2}\right)^{M} ,
\end{align*}
where in (a) we used the Schwarz inequality on the infinity operator norm, in (b) we applied the bound in~\eqref{eq:abs_prob_bound}, and in (c), where we denote by $n:= |\labelSet^{\xdim \ydim}| - |\absSet|$ the total number of transient configurations, we leveraged the upper bound on the transition probabilities (cf. Lemma~\ref{lem:bounds_transition_prob}) and the fact that $\iter \mod M < M$ which implies that the second term in (b) consists of at most $M$ factors. 
It follows that the probability of not being absorbed in $\iter$ steps, starting from an arbitrary transient state, goes to zero as $\iter \to \infty$. Then, the probability of being absorbed in finite time is $1$, regardless of $\state^0$.
\subsection{Proof of Theorem~\ref{theorem:absTime_prob_multicell}}\label{app:proof_thm_absTime_prob_multicell}

Consider the 1D-ECPM with $\numCells$ contiguous cells and tumor front at position $\indexStart$ (Figure~\ref{fig:init_state_1Dgrid_cells}). Under the forward-invasion restriction, feasible transitions divide into: 
\begin{itemize}
    \item \textit{Internal transitions}: adjacent cells invade each another through one of the $2(\numCells-1)$ internal fronts without changing the front position. The resulting configurations are thus given by $\adjConfInt := \{\Label^\prime : \maxTumorDepth{\Label^\prime} = \indexStart\}$
    \item \textit{Front-advance transition}: front cell moves to site $\indexStart+1$, leading to the (unique) configuration, represented by the singleton set $\adjConfFront := \{\Label^\prime : \maxTumorDepth{\Label^\prime} = \indexStart + 1\}$,
\end{itemize}
With $2\numCells-1$ total feasible configurations selected with equal probability, the probability of selecting the  front-advance configuration $\Label^\prime \in \adjConfFront$ is $\frac{1}{2\numCells-1}$.
Then, we compute $\probAbsTime{\indexStart+1}{\indexStart}{\iter}{\state^0}$ and the associated energy change $\Delta\E\left(\state^{\hIter}\right)$ at each step $\hIter \in \fromto{0}{\iter-1}$ of the next-site invasion process, leveraging Lemma~\ref{lem:energy_change_decomposition}. By the sparse decomposition in~\eqref{eq:deltaHJ_simplified}, the adhesion energy change for site $\indexStart+1$ yields
\begin{equation*}
    \Delta\HamJ = \sum_{w \in \Neighbour_{\indexStart+1}} \left[\energy(\Labelp{\indexStart}, \Labelp{w}) - \energy(\Labelp{\indexStart+1}, \Labelp{w})\right]
    = \energy(\labels_\numCells, \labels_\numCells) - \energy(\labelEmpty, \labels_\numCells) + \energy(\labels_\numCells, \labelEmpty) - \energy(\labelEmpty, \labelEmpty) = 0,
\end{equation*}
that cancels out based on the standard choice of the adhesion energy function in~\eqref{eq:interaction_energy_standard}.
This means that the front advance in 1D-ECPM is independent of the adhesion energy.
Instead, for the volume energy, only the front cell evolves, yielding the following change $\Delta\HamV = \lambda\Bigl[1 + 2\left(\Vol_{\labels_\numCells} - \targetVol\right)\Bigr]$.
At the first step, uniform ECM initialization in~\eqref{eq:init_cond_ecm} makes the degradation term vanish, leading to $\Delta\E\left(\state^0\right) = \lambda\Bigl[1 + 2\left(\Vol^0_{\labels_\numCells} - \targetVol\right)\Bigr]$.
For $\hIter \in \left\{1, \hdots, \iter-2\right\}$ considering a fixed tumor configuration $\LabelInit$ and evolving ECM, the associated energy is
\begin{equation}\label{eq:deltaE_h_general}
    \Delta\E\left(\state^{\hIter}\right) = \lambda\left[1 + 2\left(\Vol^{\hIter}_{\labels_\numCells} - \targetVol\right)\right] + \Kecm\left(\ecmp{\gridPoint}^{\hIter} - \ecmp{\gridPointNeigh}^{\hIter}\right),
\end{equation}
where, by Lemma~\ref{lem:asymptotic_ecm_behavior}, the ECM term $\Kecm\left(\ecmp{\gridPoint}^{\hIter} - \ecmp{\gridPointNeigh}^{\hIter}\right)$ is negative for any $\hIter$, progressively reducing the energy. 
Then, the front-advance transition probability to $\Label^{\hIter+1} \in \adjConfFront$ at next step is obtained by combining the front selection probability with the Boltzmann transition probability in~\eqref{eq:transition_prob}, giving
\begin{equation}\label{eq:prob_front_advance_step_h}
    \prob{\Label^{\hIter+1} \mid \state^{\hIter}} = \frac{1}{2\numCells-1}
    \begin{cases}
        e^{-\Delta\E^{\hIter}/\beta}
        & \text{if} \Delta\E\left(\state^{\hIter}\right) > 0, \\
        1
        & \text{otherwise}
    \end{cases}.
\end{equation}
Note that, the front advance is energetically unfavorable when $\Delta\E\left(\state^{\hIter}\right) > 0$, i.e. under ECPM parameter condition
\begin{equation}\label{eq:deltaEh_pos_condition}
    \targetVol - \Vol^{\hIter}_{\labels_\numCells} \leq \frac{1}{2} \left(\frac{\Kecm\!\left(\ecmp{\gridPoint}^{\hIter} - \ecmp{\gridPointNeigh}^{\hIter }\right)}{\lambda} + 1\right).
\end{equation}
Now, for exactly one site to advance in $\iter$ steps, the front must remain stationary for the first $\iter-1$ steps considering either internal transitions $\Label^\hIter \in \adjConfInt$ or rejected advances $\Label^\hIter \in \adjConfFront$, to then finally advance at step $\iter$. This yields
\begin{equation}\label{eq:single_site_prob_multicell}
    \probAbsTime{\indexStart+1}{\indexStart}{\iter}{\state^0} = \prob{\Label^{\iter} \mid \state^{\iter-1}}
    \prod_{\hIter=0}^{\iter-2} \psi_\hIter,
\end{equation}
where the front-advancing transition probability $\prob{\Label^{\iter} \mid \state^{\iter-1}}$ is given by~\eqref{eq:prob_front_advance_step_h}, and the terms $\psi_\hIter$ represent the probability of not advancing at step $\hIter$, defined as
\begin{equation*}
    \psi_\hIter := \frac{2(\numCells-1)}{2\numCells-1} + \frac{1 - e^{-\Delta\E\left(\state^{\hIter}\right)/\beta}}{2\numCells-1}.
\end{equation*}
Here, rearranging the terms and simplifying yields~\eqref{eq:absTime_prob_ks_multicell}.
We now extend the single-site result to multi-site progression from $\indexStart$ to $\indexStop$. With $\indexStop - \indexStart$ successive advances needed, we decompose $\iter$ total steps via ordered partitions $(\iter_0^{\partIter}, \ldots, \iter_{\indexStop-\indexStart-1}^{\partIter})$ satisfying~\eqref{eq:partition_k}, where each term specifies the timing of the $\transIter$-th transition within the $\partIter$-th partition.
For each partition, the invasion probability factors as a product of single-site terms from equation~\eqref{eq:single_site_prob_multicell}. Summing over all partitions we obtain the following invasion probability
\begin{equation*}
    \probAbsTime{\indexStop}{\indexStart}{\iter}{\state^0} = \sum_{\partIter=1}^{b(\iter,\indexStop,\indexStart)} \prod_{\transIter=i}^{\indexStop-1} \wp_\transIter^\partIter,
\end{equation*}
where $b(\iter, \indexStop, \indexStart)$ is defined as in \eqref{eq:binom_compact_def}, and $\wp_\transIter^\partIter$ is computed as in~\eqref{eq:single_site_prob_multicell} considering the step indexes referred to the $\transIter$-th transition within the $\partIter$-th partition.
This is the compact form in~\eqref{eq:absTime_prob_from_i_to_j_in_k_steps}.

%
%
\bibliographystyle{IEEEtran}
\bibliography{biblio}
\tiny

\begin{IEEEbiography}[{\includegraphics[width=0.85in,height=1.1in,clip,keepaspectratio]{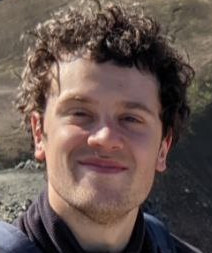}}]{Andrea Tramaloni}
  received the M.Sc.\ degree (2022) in Automation Engineering, and the Ph.D.\ in Biomedical, Electrical, and Systems Engineering (2026), from the University of Bologna, Italy.
  He was a visiting scholar at the University of Bristol, UK, in 2025.
  He is currently a Post-doc at CASY, Department of Electrical, Electronic, and Information Engineering G. Marconi of Alma Mater Studiorum Università di Bologna.
  His research interests include system biology, multi-agent systems, and control theory.
\end{IEEEbiography}

\begin{IEEEbiography}[{\includegraphics[width=0.85in,height=1.1in,clip,keepaspectratio]{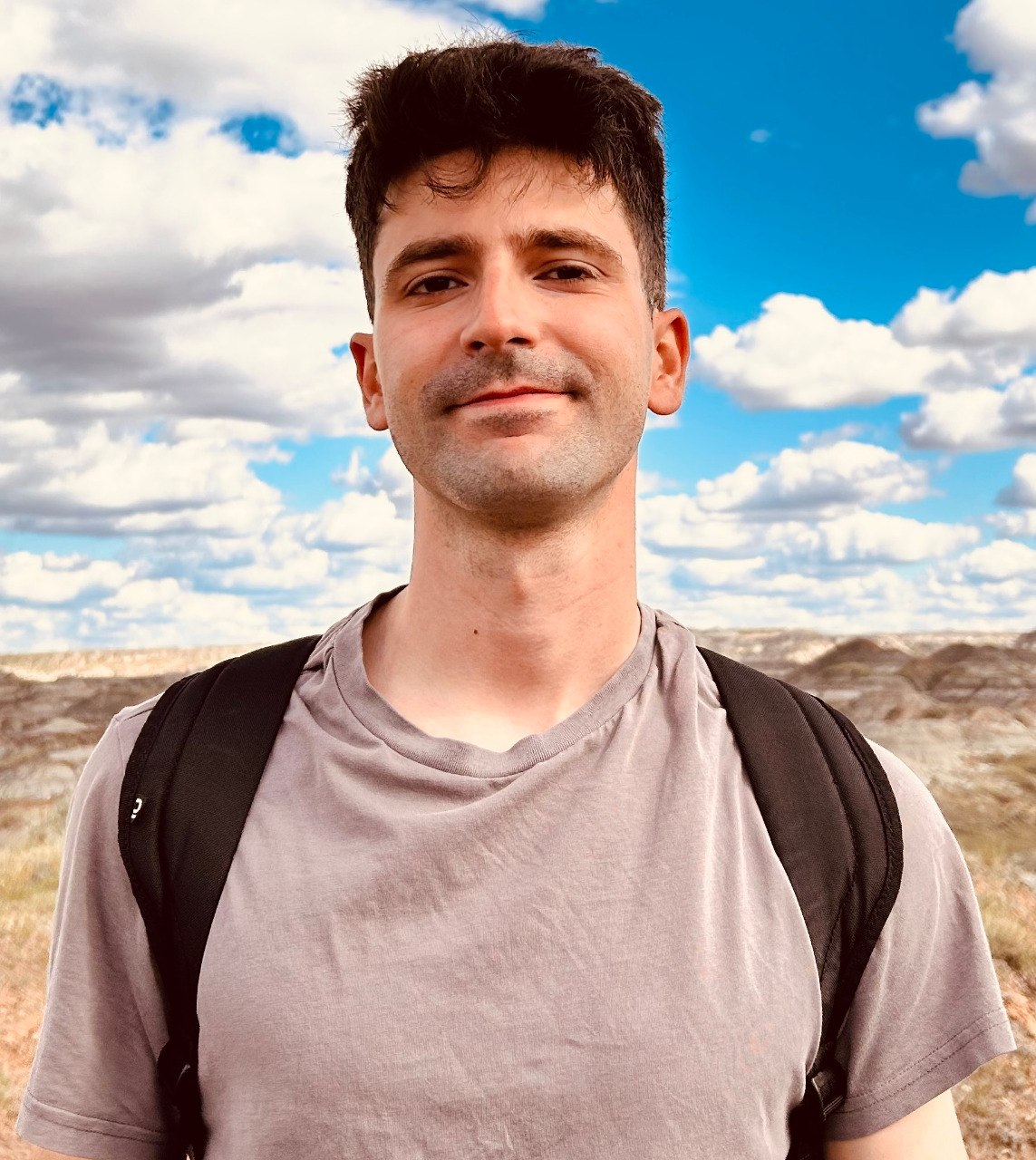}}]{Simone Baroncini} 
    received the B.Sc.\ degree (2019), the M.Sc.\ degree “summa cum laude” (2022) in Automation Engineering, and the Ph.D.\ in Biomedical, Electrical, and Systems Engineering (2026), from the University of Bologna, Italy. He was a Visiting Scholar at Queen's University, Kingston, Ontario, Canada, in~2025. He is currently a Post-doc at CASY, Department of Electrical, Electronic, and Information Engineering G. Marconi of Alma Mater Studiorum Università di Bologna. His research interests are in the intersections of reinforcement learning, control theory, and geometry.
\end{IEEEbiography}

\begin{IEEEbiography}[{\includegraphics[width=0.85in,height=1.1in,clip,keepaspectratio]{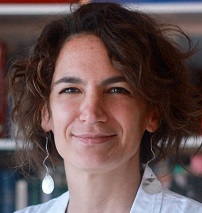}}]{Sofia Avnet} 
  received a Degree in Medical Biotechnologist, in 2000, and the Ph.D. degree, in 2005, a second Degree in cellular and molecular biology, in 2010, and the Specialization degree in pharmacology and toxicology, in 2023. During the first master's degree, she began research on bone tumors with the Kimmel Cancer Institute and the IRCCS Istituto Ortopedico Rizzoli. For the Ph.D. degree, she continued studying biological mechanisms of bone cancers with the Institute for Cancer Research, Candiolo. She also conducted research on bone pathophysiology as a
  Visiting Researcher with McGill University, in 2006, and a Research Fellow with the University of Turku, in 2007. From 2008 to 2019, she was a Biotechnologist with the Rizzoli Orthopedic Institute, focusing her research on tumor invasiveness, chemoresistance, microenvironment, and metabolism in bone cancers. Since 2019, she has been an Assistant Professor with the University of Bologna, where she develops 3D bone tissue and bone cancer models using spheroids, microfluidics, and bioprinting. Over her career, she has received several research grants as a Principal Investigator from the Italian Association of Cancer Research and the Ministry of Health. Dr. Avnet has been the President of the International Society of Cancer Metabolism (2015-2016).
\end{IEEEbiography}

\begin{IEEEbiography}[{\includegraphics[width=0.85in,height=1.1in,clip,keepaspectratio]{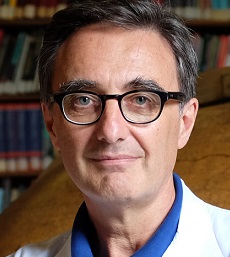}}]{Nicola Baldini} 
  received the Degree in medicine and surgery, the Certification degree in orthopaedics, and the Certification degree in oncology from the University of Bologna, Bologna, Italy, in 1981, 1987, and 1990, respectively.
  From 1988 to 1989, he was a Research Fellow with Harvard University. 
  From 2013 to 2018, he was a Visiting Professor with Kyoto Prefectural
  University of Medicine. 
  He is currently a Professor with the Department of Biomedical and Neuromotor Sciences, Alma Mater Studiorum, University of Bologna, and the Director of the Department of
  Research, Innovation and Technology, IRCCS Istituto Ortopedico Rizzoli, Bologna. 
  His research interests include the microenvironment of bone tissue under physiological and pathological conditions, bone cancers, and regenerative medicine.
  Prof. Baldini has served as the President for the Connective Tissue Oncology Society, from 1999 to 2000, the European Orthopaedic Research Society, from 2008 to 2012, and the International Society for Cancer Metabolism, from 2019 to 2020.
\end{IEEEbiography}

\begin{IEEEbiography}[{\includegraphics[width=0.85in,height=1.1in,clip,keepaspectratio]{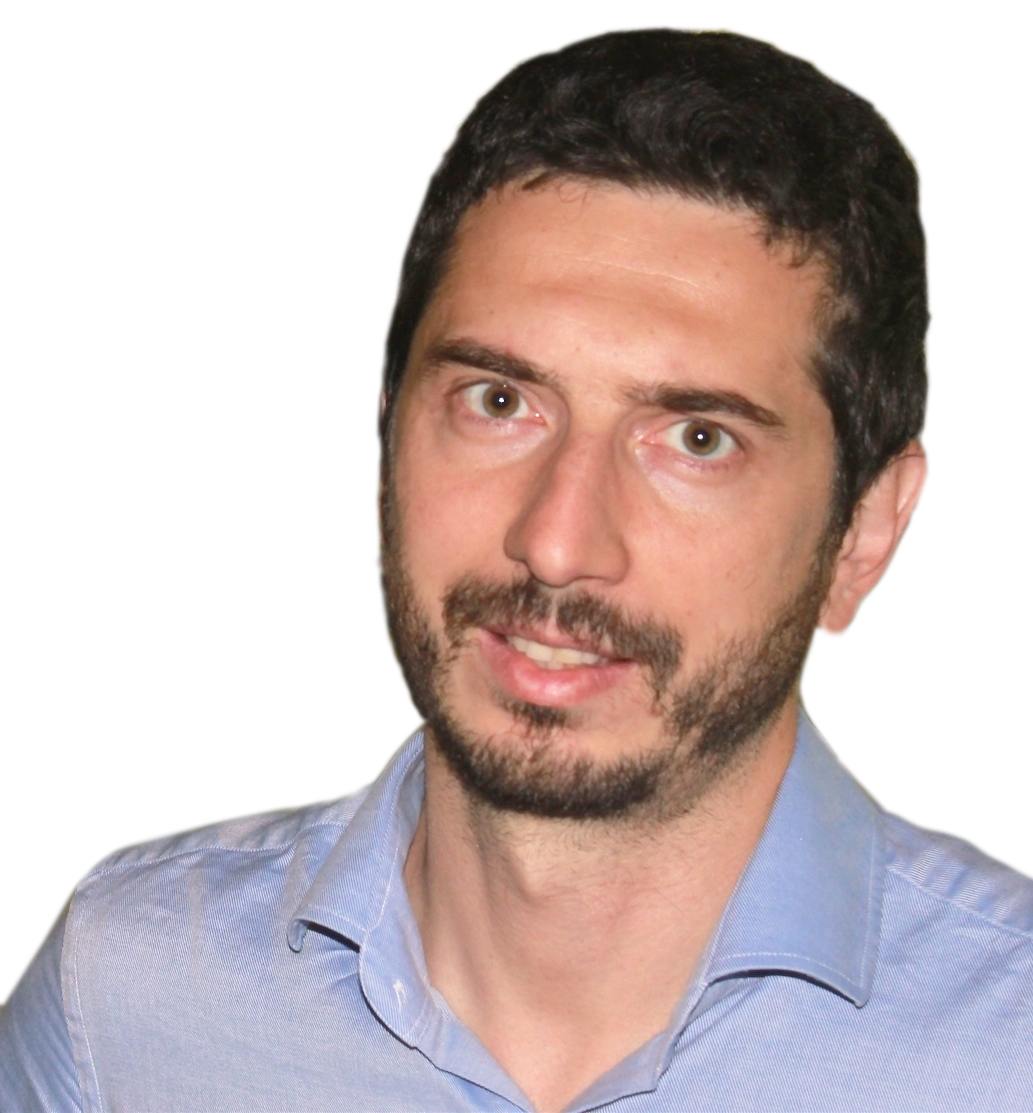}}]{Giuseppe Notarstefano} 
  is a Professor in the Department of Electrical, Electronic, and Information Engineering G. Marconi at Alma Mater Studiorum Universit\`a di Bologna. 
  He was Associate Professor (June '16 - June '18) and previously Assistant Professor, Ricercatore, (February '07 - June '16) at the Universit\`a del Salento, Lecce, Italy. 
  He received the Laurea degree “summa cum laude” in Electronics Engineering from the Universit\`a di Pisa in 2003 and the Ph.D. degree in Automation and Operation Research from the Universit\`a di Padova in 2007. 
  He has been visiting scholar at the University of Stuttgart, University of California Santa Barbara and University of Colorado Boulder. 
  His research interests include distributed optimization, cooperative control in complex networks, applied nonlinear optimal control, trajectory optimization and maneuvering of aerial and car vehicles. 
  He serves as an Associate Editor for IEEE Transactions on Automatic control, IEEE Transactions on Control Systems Technology and IEEE Control Systems Letters. 
  He is also part of the Conference Editorial Board of IEEE Control Systems Society and EUCA. 
  He is recipient of an ERC Starting Grant.
\end{IEEEbiography}

\end{document}